\documentclass[a4paper,11pt]{article}
\pdfoutput=1
\usepackage{graphicx,rotating,hyperref,slashed,amsmath,xcolor,amssymb,amsfonts,colortbl,cite}
\makeatletter
\hypersetup{colorlinks,bookmarksopen,bookmarksnumbered,
linkcolor=blus,pdfstartview=FitH,urlcolor=rossos,citecolor=verde}
\allowdisplaybreaks	 
\numberwithin{equation}{section}
\hypersetup{%
    ,urlcolor=black
    ,citecolor=black
    ,linkcolor=black
    }
\usepackage{mciteplus}

\AtBeginDocument{% optionally set colors to your liking
  \hypersetup{
    urlcolor=blue,
    citecolor=blue,
    linkcolor=black,
  }%
}

\newcommand{\troisj}[6]{\left(\begin{array}{ccc}
      #1 & #2 & #3 \\
      #4 & #5 & #6\end{array}\right)}

\def\beq{\begin{equation}}
\def\eeq{\end{equation}}

\newcommand{\bv}{{\bf{v}}}

\newcommand{\bn}{{\bf{n}}}

\newcommand\ees{\end{eqnarray}}
\newcommand\bees{\begin{eqnarray}}

\def\bea{\begin{eqnarray}}
\def\eea{\end{eqnarray}}

\def\dd{{\rm d}}

\def\nn{\nonumber}

\def\0{{\boldsymbol 0}}

\def\lsim{\mathrel{\rlap{\lower3pt\hbox{\hskip0pt$\sim$}}
   \raise1pt\hbox{$<$}}}         %less than or approx. symbol
\def\gsim{\mathrel{\rlap{\lower4pt\hbox{\hskip1pt$\sim$}}
   \raise1pt\hbox{$>$}}}         %greater than or approx. symbol

 \newcommand{\sfootnote}[1]{}
\definecolor{bluc}{cmyk}{1,1,0,0.1}
\definecolor{rossoCP3}{cmyk}{0,.88,.77,.40}
\definecolor{rosso}{cmyk}{0,1,1,0.4}
\definecolor{rossos}{cmyk}{0,1,1,0.55}
\definecolor{rossoc}{cmyk}{0,1,1,0.2}
\definecolor{verdes}{cmyk}{0.92,0,0.59,0.4}

\hypersetup{colorlinks, bookmarksopen, bookmarksnumbered,
citecolor=verdes, linkcolor=bluc, pdfstartview=FitH, urlcolor=rossos}

\usepackage{multicol}
\usepackage{color}
\definecolor{rosso}{cmyk}{0,1,1,0.4}
\definecolor{rossos}{cmyk}{0,1,1,0.55}
\definecolor{rossoc}{cmyk}{0,1,1,0.2}
\definecolor{blu}{cmyk}{1,1,0,0.3}
\definecolor{blus}{cmyk}{1,1,0,0.6}
\definecolor{bluc}{cmyk}{1,1,0,0.1}
\definecolor{verde}{cmyk}{0.92,0,0.59,0.25}
\definecolor{verdec}{cmyk}{0.92,0,0.59,0.15}
\definecolor{verdes}{cmyk}{0.92,0,0.59,0.4}

\skip\@mpfootins = \skip\footins
\def\circa#1{\,\raise.3ex\hbox{$#1$\kern-.75em\lower1ex\hbox{$\sim$}}\,}

\newcommand{\be}{\begin{equation}}
\newcommand{\ee}{\end{equation}}
\newfam\rsfsfam
\def\mathscr#1{{\fam\rsfsfam\relax#1}}

\def\circa#1{\,\raise.3ex\hbox{$#1$\kern-.75em\lower1ex\hbox{$\sim$}}\,}
\makeatletter

\def\hhref#1{\href{http://arxiv.org/abs/#1}{arXiv:#1}} % in bibliography

\newcommand{\doi}[1]{\href{http://dx.doi.org/#1}{[doi]}}

\def\hhref#1{\href{http://arxiv.org/abs/#1}{arXiv:#1}} 
 
\def\art{\@ifnextchar[{\eart}{\oart}}
\def\eart[#1]#2#3#4#5#6{{\rm #2}, {\em #3 \bf #4} {\rm (#6) #5} ({\em #1})}

\def\article{\@ifnextchar[{\earticle}{\oarticle}}
\def\oarticle#1#2#3#4#5#6{{\rm #1}, {\em ``#6''}, {\rm #2 #3 (#5) #4}}
\def\earticle[#1]#2#3#4#5#6#7{{\rm #2}, {\em ``#7''}, {\rm #3 #4 (#6) #5}  [\hhref{#1}]}
\def\hepart[#1]#2{{\rm #2, \em#1}}
\def\heparticle[#1]#2#3{#2, {\em ``#3''} [\hhref{#1}]}

\newcounter{alphaequation}[equation]
\def\thealphaequation{\theequation\hbox to
0.6em{\hfil\alph{alphaequation}\hfil}}
\def\eqnsystem#1{
\def\@eqnnum{{\rm (\thealphaequation)}}
\def\@@eqncr{\let\@tempa\relax \ifcase\@eqcnt \def\@tempa{& & &} \or
  \def\@tempa{& &}\or \def\@tempa{&}\fi\@tempa
  \if@eqnsw\@eqnnum\refstepcounter{alphaequation}\fi
\global\@eqnswtrue\global\@eqcnt=0\cr}
\refstepcounter{equation} \let\@currentlabel\theequation \def\@tempb{#1}
\ifx\@tempb\empty\else\label{#1}\fi
\refstepcounter{alphaequation}
\let\@currentlabel\thealphaequation
\global\@eqnswtrue\global\@eqcnt=0 \tabskip\@centering\let\\=\@eqncr
$$\halign to \displaywidth\bgroup \@eqnsel\hskip\@centering
$\displaystyle\tabskip\z@{##}$&\global\@eqcnt\@ne
\hskip2\arraycolsep\hfil${##}$\hfil& \global\@eqcnt\tw@\hskip2\arraycolsep
$\displaystyle\tabskip\z@{##}$\hfil
\tabskip\@centering&\llap{##}\tabskip\z@\cr}
\def\endeqnsystem{\@@eqncr\egroup$$\global\@ignoretrue} \makeatother

\definecolor{fiorentina}{rgb}{.5,0,.5}

\begin{document}
%\preprint{ET-0465A-21}

%ET-0465A-21
\setcounter{page}{1} \baselineskip=15.5pt \thispagestyle{empty}

%\vskip-1.5cm
\hskip13.5cm  ET-0003A-22 

%\smallskip\
 
\vspace{0.8cm}
\begin{center}

{\fontsize{19}{28}\selectfont  \sffamily \bfseries {Doppler boosting  the stochastic gravitational wave background}}
%Constraining the SGWB  spectral shape 
%\vspace{0.7 em}\\
%with Doppler induced anisotropies }}%Doppler boosting  the SGWB

\end{center}

\vspace{0.2cm}

\begin{center}
{\fontsize{13}{30}\selectfont Giulia Cusin$^{1,2}$, Gianmassimo Tasinato$^{3}$ } 
\end{center}

\begin{center}

\vskip 8pt
\textsl{$^{1}$ Department of Theoretical Physics, University of Geneva, 24 quai Ernest-Ansermet, Geneva, Switzerland}
\\
\textsl{$^{2}$Sorbonne Université, CNRS, UMR 7095, Institut d'Astrophysique de Paris, 75014 Paris, France}
\\
\textsl{$^{3}$ Physics Department, Swansea University, SA28PP, United Kingdom}\\
\vskip 7pt

\end{center}

\smallskip
\begin{abstract}
\noindent
One of the guaranteed features of the stochastic gravitational wave background (SGWB)
  is the presence of  Doppler anisotropies
induced by the motion of the detector with respect to the rest frame of the SGWB
source.  We point out that kinematic effects can be  amplified
 if the SGWB   is characterised by large tilts in its  spectrum as a function of frequency, or
 by sizeable intrinsic anisotropies.   Hence we examine the possibility to use Doppler  effects as
   complementary probes of the SGWB frequency profile. 
For this purpose
we work in multipole space, and we study the effect of kinematic modulation and aberration on the GW energy density parameter and on its angular power spectrum. We develop a Fisher forecast analysis and we discuss prospects   for constraining   parameters controlling kinematically induced anisotropies  with future detector networks.  
As a case study, we apply our framework to a background component with constant slope in frequency, potentially detectable by a network of future ground-based interferometers. For this specific example, we show that a measurement of kinematic anisotropies with a network of Einstein Telescope and Cosmic Explorer will allow us to constrain the spectral shape with a precision of about 16\%.  {We also show that, if a reconstruction of the spectral shape is done via other methods, e.g. frequency binning, a study of kinematic anisotropies can allow one to constrain our peculiar velocity with respect to the CMB frame with a precision of $30\%$. }
Finally, we identify cosmological and astrophysical scenarios   where
  kinematic effects are enhanced  in frequency ranges probed by current and future GW experiments. 
\end{abstract}

\newpage

\section{Introduction}

The detection and  characterization of a stochastic gravitational wave background (SGWB) is one of the next goals for gravitational wave (GW) science. 
 A SGWB can have astrophysical or cosmological
 origin, or both: see  \cite{Regimbau:2011rp,Caprini:2018mtu,Maggiore:2018sht} for  recent reviews.
Currently, upper limits are set on the amplitude \cite{Abbott:2021xxi} of the SGWB at ground-based interferometer frequencies, and  on parameters characterising the {first
few multipoles of its associated angular power spectrum}
%first 
% of its properties 
  \cite{TheLIGOScientific:2016xzw,Abbott:2021jel}.  Recent  tantalising hints of a signal in the nano-Hertz regime are
 discussed  in  \cite{Arzoumanian:2020vkk}. 

In view of future conclusive detections, it is essential to have the best possible   theoretical understanding  of  
the properties of a SGWB.
 Among  its guaranteed features  is the presence of  Doppler anisotropies
induced by the motion of the detector with respect to the SGWB rest frame. In fact, the study of kinematic anisotropies {in the context of stochastic backgrounds of electromagnetic radiation} is an already well-developed research topic.
%  already
 % well
  % been detected and/or  theoretically well 
%studied.
  Doppler-induced 
anisotropies in the spectrum of CMB fluctuations have been measured since decades
 \cite{Smoot:1977bs,Kogut:1993ag,Bennett:2003bz,Aghanim:2013suk},
 % where
%they provide the largest source of anisotropy. They
and their   analysis 
provides interesting cosmological information \cite{Henry:1969im,Peebles:1968zz,Gorski:1990ua,CvL02,Menzies:2004vr,BR06,Kosowsky:2010jm,Amendola:2010ty,Mukherjee:2013zbi}. Kinematic effects are also studied for other cosmological observables, see , see e.g. the
%
%
%\cite{Gorski:1990ua,Kogut:1993ag,Aghanim:2013suk,Amendola:2010ty}
%\item Kinematic anisotropies in other probes: 
 dipole in luminosity distance of SN counts \cite{Bonvin:2006en}, or the kinematic  dipole anisotropies  in 
 galaxy  number counts, see e.g.
  \cite{10.1093/mnras/206.2.377,Maartens:2017qoa,Pant:2018smd}.
  
  \smallskip
  The aim of this work is to  analyse 
   kinematic anisotropies in the SGWB, and investigate
   their physical consequences. In particular, we  
   show that their properties 
  can provide important information on the spectral dependence of the SGWB:  
   
   \begin{itemize}
   \item 
   In  section \ref{sec-trsci} we compute how  the SGWB density parameter $\Omega_{\rm GW}$  transforms under a Doppler boost  connecting the observer frame to a  frame at rest with the SGWB. The boost  leads to the generation of  kinematic anisotropies from  the rest-frame monopole, and  to the modulation and aberration of existing   rest-frame anisotropies.   Kinematic effects are  amplified
 if the SGWB   is characterised by sizeable tilts in its  spectrum as a function of frequency, or
 by  intrinsic anisotropies. 
 Consequently, Doppler  effects offer
   complementary probes of the SGWB frequency profile, as well as of its  rest-frame anisotropies. 
 \item
 In section  \ref{sec-knees} we identify set-ups where
  kinematic effects can be amplified  in frequency ranges probed by GW experiments, thanks to transient enhancements of 
the tilt of the SGWB spectrum. For cosmological background components, such amplifications
can occur when tensor modes are sourced
at second order by  scalar spectra with pronounced peaks, as in inflationary models producing  primordial black holes. For an astrophysical background, scenarios with a rapid after merger drop of the density parameter can potentially lead to a local enhancement of kinematic anisotropies. 
\item In 
 section \ref{sec-prospects}  we  discuss     applications
of our  results, elaborating prospects of detection of kinematically-induced effects. 
We  investigate    constraints on the slope of the SGWB spectrum associated with 
 measurements of boost induced anisotropies. 
  We develop a multipolar decomposition of the  SGWB spectrum including Doppler
  effects. We  show how the latter depend on the tilt of the spectrum, 
  and how they 
   affect correlation functions. We
develop Fisher forecasts for
  computing the signal-to-noise ratio relative to parameters controlling    kinematically induced anisotropies. We 
  then
  apply our general formulas to  
  the idealised case of
  an astrophysical background with constant  slope in frequency,  detectable by a network of future  interferometers. 
  For this specific example, we show that a measurement 
  of kinematic anisotropies  allows  us to constrain the spectral shape of the SGWB with a precision of about $16\%$. {We also show that, if a reconstruction of the spectral shape is done via other methods, e.g. frequency binning, a study of kinematic anisotropies can allow one to constrain our peculiar velocity with respect to the CMB frame with a precision of $30\%$.}
   \item  We present our conclusions and future directions of our study in section \ref{sec-outlook}, which is followed by a
    technical appendix \ref{lm}.
\end{itemize}

\section{Transforming the GW density parameter under boosts}
\label{sec-trsci}

In this section we discuss how  the GW density parameter $\Omega_{\rm GW}$
transforms
under a Doppler boost   connecting two frames  moving with relative velocity ${\bf v}$.
In subsection \ref{sec-genfo1}
 we obtain
  the general formulas for boost transformations.
In subsection \ref{sec-example1}
 we apply them to a case where  the GW density parameter $\Omega_{\rm GW}$ is isotropic in the rest frame, and we demonstrate that  kinematic anisotropies are induced  by the boost\footnote{Part of the contents of these
 two subsections can be found in \cite{Bartolo:2022pez}.}.
In subsection \ref{sec-example2}  we discuss a more general scenario where the GW density parameter $\Omega_{\rm GW}$ is 
anisotropic in the rest frame. We show how kinematic effects, besides
inducing new anisotropies,  lead to  modulation  and aberration of existing ones. The size of the
Doppler-induced effects are enhanced if the SGWB spectrum has large tilts in frequency. This 
property suggests an independent way for probing the frequency  dependence of the spectrum. 
 See also \cite{Jenkins:2018lvb} for an analysis of the kinematic dipole of the SGWB, with applications
 to the case of cosmic string sources.

\subsection{The general formulas}
\label{sec-genfo1}
Let us consider two sets of observers in an unperturbed FLRW universe.
%\begin{enumerate}
%\item
 The first observer, called   $\mathcal{S}'$,  is comoving with the SGWB rest frame.
 %  \textcolor{magenta}{(which we identify with the CMB rest frame)}.
 %\item
  The second one,  $\mathcal{S}$,  moves with constant velocity ${\bf v}$ with respect to
    $\mathcal{S}'$.
%  \end{enumerate}
Primes indicate rest-frame quantities;     vectors  with a hat are  
unit vectors.  The 
boost  connects the
 SGWB density parameters    in the two frames 
 %
  %in the rest frame 
  $\mathcal{S}'$
  % (rest frame)
  and
  % to the one in the moving one 
   $\mathcal{S}$.  %(moving frame).
 We  identify two qualitatively different  effects of  the observer motion  on the SGWB sky map\footnote{While we  investigate these effects in the case of the SGWB, it is worth to point out  that they are already 
%These effects are 
well-known in the CMB literature.}: 
\begin{enumerate}
\item 
The
  generation of higher multipole  anisotropies from lower multipole ones.  
  \item  The modulation and aberration of the intensity of  existing anisotropies. This leads to
      a remapping of the intensity map/Stokes parameters on the sky. 
\end{enumerate}
 We use the approach of 
 \cite{mckinley} 
   to determine how quantities change from one frame to the other (see also \cite{Kosowsky:2010jm,Peebles:1968zz,Landau:1987gn}).
We denote with  $f'$ the frequency  of the GW in the  SGWB rest frame, and with $\hat \bn'$
the unit vector denoting 
its direction. The frequency $f$ in the system in motion  is related to $f'$ by a Lorentz
transformation reading
% a boost the frequency changes as
\be\label{Eshift}
f\,=\,\frac{\sqrt{1-\beta^2}}{1- \beta\,\xi}f'\,,
\ee
where $\bv=\beta \hat{\bv}$ denotes the relative velocity of the two frames: we have $\beta=v$ in units with $c=1$. The unit vector 
$\hat{\bv}$ %is a unit vector 
 corresponds to  the direction of the relative motion between the two frames. We  introduce  the convenient quantity
\be
\label{defmu}
\xi\,=\, \hat\bn \cdot\hat   \bv\,,
\ee
parametrising the relative angle between  $\hat \bn$ and $\hat \bv$.  
We re-express  equation (\ref{Eshift}) as
 %in the more compact form 
\bea
\label{defcaldA}
f&=&{\cal D}\,f' \label{rela1}\,,
\eea
with
%where we define
\be
\label{defcald}
{\cal D}\,=\,\frac{\sqrt{1-\beta^2}}{1- \beta \,\xi}\,. 
\ee
The directions of GW propagation   in  the two  frames 
$\mathcal{S}'$
  % (rest frame)
  and
  % to the one in the moving one 
   $\mathcal{S}$   % relative to the velocity $\hat \bv$ 
 are related by the aberration equation  of special relativity \cite{Aghanim:2013suk}
% 
%\bea\label{aberration}
%\hat \bn  \hat{\bv}&=&\frac{\hat \bn'  \hat{\bv}+\beta}{1+\beta\,\hat \bn' \hat \bv}\,.
%\eea
%
%Using special relativistic aberration formulas, one finds that
 %the unit vector $\hat \bn'$ is expressed in terms of the vector $\hat \bn$ as \cite{Aghanim:2013suk}
\be \label{trnp1}
\hat \bn'\,=\,\frac{\hat \bn+\hat \bv \left[ \left(\gamma-1 \right)\xi -\gamma \beta\right]}{\gamma \left( 1-\beta \xi\right)}\,,
\ee
with
\bea
\label{defga}
\gamma&=& \frac{1}{\sqrt{1-\beta^2}}\,.
%\\
%\xi&=& \hat \bn \hat \bv 
\eea

In order  to 
 compute  how the GW energy density changes under  boosts, we make use  of the GW distribution function, denoted here with
 $\Delta'( f', \hat \bn' )$. We assume  it only depends
 on the frequency $f'$ and  on the GW direction $\hat \bn'$
 in the SGWB rest frame. 
 We  express  the number of gravitons for unit of phase space  in the rest-frame ${\cal S}'$ as:
 \be
 \label{dNgrnumb}
d N' \,=\,\Delta'(f', \hat \bn'   )\,f'^2\,d f'\,d^2 \hat \bn'\,d V'\,,
\ee
where  $dV'$ corresponds to the infinitesimal volume containing
 gravitons with propagation vector $\hat \bn'$ in the element of measure $ d f'\, d^2 \hat \bn'$.
   %It is not difficult to prove that 
  The combination  $f'^2\,d f'\,d^2 \hat \bn'\,d V'$ is invariant
  under boosts. In fact,  the relations $f'\,=\,{\cal D}^{-1}\,f$, $d^2\hat  \bn'\,=\,{\cal D}^{2}\, d^2 \hat \bn$,
  $d V'\,=\,{\cal D}\,d V$ hold (see  \cite{mckinley,Kosowsky:2010jm}).  On the other hand, the number of gravitons \eqref{dNgrnumb} is independent
of the frame, and $dN '\,=\,d N$. Hence 
 \cite{Landau:1987gn}
  \be
  \label{samdi1}
\Delta'( f', \hat \bn'   )\,=\,\Delta(f, \hat \bn   )\,.
\ee
The GW distribution function $\Delta$ can be used to define the energy density of GW in the rest frame as
energy per unit volume and unit solid angle:
\bea
d \rho'_{\rm GW}(f', \hat \bn'  )&=& \frac{f'\,d N' }{d^2 \hat \bn'\,d V'}
%f(\omega')\,\omega'^3\,d \omega'
\,=\,\Delta'(f', \hat \bn'  )\,f'^3\,d f'\,.
\eea
%This definition allows us 
We then express the GW density parameter $\Omega_{\rm GW}'(\omega', \hat \bn'  \hat \bv)$
in the  rest frame ${\cal S}'$ as
\begin{eqnarray}
\label{genexpom}
\Omega'_{\rm GW}(f', \hat \bn'  )\equiv\frac{1}{\rho_c}\,\frac{d  \rho'_{\rm GW}}{d \ln f'}&=&\frac{3\pi\,f'^4}{2\,H_0^2\,M_{\rm Pl}^2}\,\Delta'(f',  \hat \bn' )\,.
%\\
%&=&\frac{\omega^4}{\omega'^4}\,\Omega'_{\rm GW}( \omega')
\end{eqnarray}
Using eq \eqref{samdi1}, we find the equality
\be
\Omega_{\rm GW}(f, \hat \bn  )\,=\,\left( \frac{f}{f'} \right)^4 \Omega'_{\rm GW}(f', \hat \bn' )\,.
\ee
%Using special relativistic aberration formulas, one finds that
 %the unit vector $\hat \bn'$ is expressed in terms of the vector $\hat \bn$ as \cite{Aghanim:2013suk}
%\be \label{trnp1}
%\hat \bn'\,=\,\frac{\hat \bn+\hat \bv \left[ \left(\gamma-1 \right)\xi -\gamma \beta\right]}{\gamma \left( 1-\beta \xi\right)}\,,
%\ee
%with
%\bea
%\label{defga}
%\gamma&=& 1/\sqrt{1-\beta^2}\,.
%\\
%\xi&=& \hat \bn \hat \bv 
%\eea
%Eq \eqref{trnp1} can be expanded at linear order in $\beta$ as  \cite{Aghanim:2013suk} (see
%also \cite{Challinor:2002zh})
%\bea
%\hat \bn'&\simeq&\hat \bn-\beta \left( \hat \bv-\xi\,\hat \bn\right)
%\\
%&=&\hat \bn-\bold{\nabla} \left( \beta \xi\right)
%\eea
%where $\bold{\nabla}$ is the  gradient, and we used the identity $\bold{\nabla} (\beta \,\hat \bv \cdot \hat \bn)
%\,=\,\beta \,\hat \bv\cdot \bold{\nabla}\,\, \hat \bn
%\,=\,\beta \left( \hat \bv - \hat \bv\cdot \,\,\hat \bn \right)$.
%We will use this relation  in what follows.

Collecting the  results so far, 
%relating the quantities $f$ and $\hat \bn$ in the two frames (eqs \eqref{defcaldA} and \eqref{trnp1}), 
 we find that the GW density parameter in the moving frame ${\cal S}$ is related with the corresponding quantity in the frame  ${\cal S'}$ at rest through the
 formula  
\be
\label{genexpom4}
\boxed{\Omega_{\rm GW}(f, \hat \bn )\,=\,{\cal D}^4\,\, \Omega'_{\rm GW}\left({\cal D}^{-1}\,f, 
\frac{\hat \bn+\hat \bv \left[ \left(\gamma-1 \right)\xi -\gamma \beta\right]}{\gamma \left( 1-\beta \xi\right)}
%\frac{\bn \hat{\bv}-\beta}{1-\beta\,\bn \hat \bv}
%\bn'  \hat \bv )
\right)}
\ee
%where in the second line we used \eqref{vbt1}.
 with ${\cal D}$, $\gamma$ and $\xi$ given respectively in eqs \eqref{defcald}, \eqref{defga}, \eqref{defmu}.
 The previous formula is completely general and valid for any values of $0\le\beta\le1$. On the other hand, 
 the parameter $\beta$ is usually small: 
 for example, for cosmological backgrounds,  CMB suggests that 
 %\textcolor{magenta}{if we identify ${\cal S'}$ with the CMB rest frame,
  $\beta\simeq 1.23 \times 10^{-3}$. Under the assumption of small $\beta$, 
  we Taylor expand  eq \eqref{genexpom4} in $\beta$,  
  and analyse two cases.
 
\subsection{First example: the SGWB is isotropic in the rest frame}
\label{sec-example1}

We assume that the  GW density parameter in the rest frame ${\cal S}'$ is isotropic, and independent
of $\hat \bn'$: $\Omega'_{\rm GW}\,=\,\Omega'_{\rm GW}(f')$.  It is then straightforward to expand
eq \eqref{genexpom4}
 up to the quadrupole (and also beyond if needed, see section \ref{sec-prospects}). 
Taking  the notation from CMB physics, we introduce the  tilts of the SGWB spectrum as
\bea
\label{defno}
n_{\Omega}(f)&=&\frac{d\,\ln \Omega'_{\rm GW}(  f)}{d\,\ln f}\,,
\\
\label{defao}
\alpha_{\Omega}(f)&=&
\frac{d\,n_{\Omega}(f)}{d\,\ln f}\,.
\eea
%a
%\bea
%n_{\Omega}&=&\frac{d\,\ln \Omega'_{\rm GW}(  \omega')}{d\,\ln \omega'}
%\hskip1cm,\hskip1cm\alpha_\Omega\,=\,
%\frac{d\,n_{\Omega}}{d\,\ln \omega'}
%\\
%\beta_\Omega&=&
%\frac{d\,\alpha_\Omega}{d\,\ln \omega}
%\hskip1cm,\hskip1cm\gamma_\Omega\,=\,
%\frac{d\,\beta}{d\,\ln \omega}
%\eea
These spectral tilts  play an important role in our analysis. 
Expanding \eqref{genexpom4}  in powers of 
% to
 %order 
 $\beta$,
 % we learn that  anisotropies of GW energy density  are generated
 %are created at all orders in a multipole expansion 
% from the monopole of an isotropic SGWB in its rest frame.
 and limiting the expansion to order
 $\beta^2$ -- in fact  we are assuming  that $\beta$ is small -- we find that 
 the GW density parameter in the moving frame ${\cal S}$ receives
   a kinematic modulation of the monopole. Moreover,   a kinematic dipole  and 
 a kinematic quadrupole are generated by  boost effects:
% the quadrupole ($\ell=2$), we get (we keep only terms propto $\beta^2$) 
\bea
\Omega_{\rm GW}(f, \hat \bn)&=&\Omega'_{\rm GW}(f)
\left[ 1+M(f)+
 \xi\,D(f)+
 \left( 
 \xi^2-\frac13
 %(\hat \bn \,\hat \bv)^2-\frac13  
  \right)
 \,Q(f)
 \right]\,,
  \label{genexpom4a}
\eea
where remember we define  $\xi\,=\,\hat \bn \cdot \hat \bv$. The frequency-dependent coefficients
%with
\bea
\label{monanis}
M(f)&=&\frac{\beta^2}{6} \left( 8+n_\Omega \left( n_\Omega-6\right)
+\alpha_\Omega
\right)\,,
\\
\label{dipanis}
D(f)&=& \beta \left(4-n_\Omega\right)\,,
\\
\label{quapanis}
Q(f)&=&\beta^2\left(10-\frac{9 n_\Omega}{2} +\frac{n_\Omega^2}{2}+\frac{\alpha_\Omega}{2}\right)\,,
\eea
indicate respectively the monopole, dipole, quadrupole  boost contributions; from now on we
understand for simplicity the frequency dependence of the spectral tilts. 

The quantities within  brackets in the expressions  \eqref{monanis}, \eqref{dipanis},  \eqref{quapanis}
depend on numerical coefficients, as well as  on the 
spectral tilts $n_{\Omega}$ and $\alpha_\Omega$, as 
 defined in eqs \eqref{defno}, \eqref{defao}.   Notice that when the spectral tilts are of at least of order one, they give a sizeable contributions to the kinetically induced effects of eqs. \eqref{monanis}-\eqref{quapanis}.
 % or larger, 
%When
%the latter are at least of order one, they became of the same order of the former.  
  Hence the slope of the spectrum influences 
  the kinematic anisotropies through $n_\Omega$, $\alpha_\Omega$: in scenarios
where these quantities are large, boost effects are amplified,
and can be used to probe the slope of the spectrum. 
%see  next
%sections.
% \ref{sec-knees}.
%
%\bea
%\frac{\Omega_{\rm GW}(\omega)}{\Omega'_{\rm GW}(\omega)
%}&=&\left[1+\frac{\beta^2}{6} \left[ 8+n_\Omega \left( n_\Omega-6\right)
%+\alpha_\Omega
%\right] \right]
%\nonumber
%\\
%&+&
%\beta
 %(\hat \bn \, \hat \bv)\left[
 %4-n_\Omega \right]
 %% - v^2 \left( 
 %%8-4 n_\Omega +\frac{n_\Omega^2}{2}+\frac{\alpha_\Omega}{2}
 %%\right)\right]
%%\left[1+(4-n_{\Omega}) \, (\bn \,\bv)\right]
%%\\
%%\Omega_{\rm GW}( \tau, \omega)&=&{\cal D}^{4}\,\Omega_{\rm GW}({\cal D}^{-1} \tau, \,{\cal D}\,\omega)
%\nonumber
%\\
%&+&\beta^2
 %\left( (\hat \bn \,\hat \bv)^2-\frac13   \right)\times\left[ 
%10-\frac{9 n_\Omega}{2} +\frac{n_\Omega^2}{2}+\frac{\alpha_\Omega}{2}
% %+\frac{v^2}{2} \left( \right)
 %\right]
 %\label{genexpom4a}
%\eea
%Each of the contributions within squared parenthesis in eq \eqref{genexpom4a} depend only on
%frequency. 

The monopole contribution $M(f)$ receives a modulation of its intensity at order $\beta^2$ in the expansion. See
\cite{Kamionkowski:2002nd}  for a study of this effect in the context of the CMB, {
and more in general section 2 of \cite{Aghanim:2013suk} for comparing our formulas to their
analog in a CMB context, with the frequency-dependence of the CMB intensity  
taking the place of what for us is the frequency-dependence of $\Omega_{\rm GW}$}. The dipole
contribution \eqref{dipanis} is the only one starting already at order $\beta^1$ in the expansion, {and it is typically the largest boost-induced modulation effect.} 
%and its size
%is typically the dominant one in an expansion in powers of $\beta$.

%
%, while the second and third line
%is creation of anisotropies from the monopole. 

While our expression  \eqref{genexpom4a} is built in terms of  combinations of $\xi\,=\,\hat \bn \cdot \hat \bv$,
it is also straightforward to convert it in spherical harmonics. We choose  for simplicity $\hat \bv$  along the $z$-direction, 
% (more general choices are related to this by a rotation),
and parameterize $\hat \bn\,=\,{\left( \sin \theta\,\cos \varphi,\,\sin \theta  \sin \varphi,\,\cos \theta \right)}$. Then
we can rewrite  \eqref{genexpom4a}  in terms of spherical harmonics $Y_{\ell m}(\theta,\varphi)$ as
\bea
\Omega_{\rm GW}(f, \hat \bn)&=&\sqrt{4\,\pi}\,\Omega'_{\rm GW}(f)
\left[ \left(1+M(f)\right)\,Y_{00}(\theta,\varphi)+
\frac{D(f)}{\sqrt{3}}\,{Y_{10}(\theta,\varphi)}+
 %\left( (\hat \bn \,\hat \bv)^2-\frac13   \right)
 %\,
 \frac{2\,Q(f)}{\sqrt{45}}\,Y_{20}(\theta,\varphi)
% Q(f)\,{\frac{2\,Y_{20}(\theta,\varphi)}{\sqrt{45}}}
 \right]\,.
 \nonumber\\
  \label{genexpom4b}
\eea
%
%
%\begin{itemize}
%\item The monopole $1\,=\,\sqrt{4\,\pi}\,Y_{00}(\theta,\varphi)$
%\item The dipole $\hat \bn \, \hat \bv\,=\,\sqrt{\frac{4 \pi}{3}}\,Y_{10}(\theta,\varphi)$
%\item The quadrupole $ \left( (\hat \bn \,\hat \bv)^2-\frac13   \right)\,=\,\sqrt{\frac{16 \pi}{45}}\,Y_{20}(\theta,\varphi)$
%\end{itemize}
Choosing $\hat \bv$ in the $z$-direction implies that only the $m=0$ harmonics are induced by the boosts -- 
more general choices induce other harmonics as well, and are related to the previous formula by a spatial rotation
(see section \ref{sec-prospects}, especially footnote \ref{ftn-rot}.). 

%%%%%%%%%%%%%%%%%%%%%%%%%%%%%%%%%%%%%%%%%%%%%%%%%%
\subsection{Second example: the SGWB is anisotropic in the rest frame}
\label{sec-example2}
%%%%%%%%%%%%%%%%%%%%%%%%%%%%%%%%%%%%%%%%%%%%%%%%%%

Doppler boosts cause aberration effects that change the map distribution 
of rest-frame anisotropies in the sky. We investigate this effect in 
%
%\smallskip
%\noindent
%{\bf Case 2:}
%For 
our second example,
 where we   do not assume that $\Omega_{\rm GW}'$ is isotropic in the rest frame ${\cal S}'$. For simplicity we
 assume  a factorisable Ansatz \cite{Allen:1996gp}: 
 \be
 \label{factcon1}
 \Omega_{\rm GW}'(f',\, \hat \bn' )\,=\,\Omega'(f') \Phi'(  \hat\bn'   )\,.
 \ee
 We now derive  the resulting $\Omega_{\rm GW}(f,\,\hat \bn )$ in the moving frame ${\cal S}$. We will find  that the quantity
  $\Omega_{\rm GW}(f,\,\hat \bn )$ in the moving frame  ${\cal S}$ does   not obey  any more a  factorisable Ansatz as in eq \eqref{factcon1}. 
  In fact, using eq \eqref{genexpom4} 
  we  express $\Omega_{\rm GW}$  as (recall $\xi\,=\,\hat \bv \hat \bn$)
 \be
\label{genexpom5a}
\Omega_{\rm GW}(f,  \hat \bn )\,=\, 
\left[{\cal D}^4  \Omega' \left({\cal D}^{-1}\,f
\right) \right]\times
\left[  
\Phi'\left(
\frac{\hat \bn+\hat \bv \left[ \left(\gamma-1 \right)\xi -\gamma \beta\right]}{\gamma \left( 1-\beta \xi\right)}
\right)\right]\,.
\ee
Expanding up to first order $\beta^1$, we get
 \be
\label{genexpom6a}
\Omega_{\rm GW}(f,  \hat \bn )\,=\, 
 \Omega_{\rm GW}'(f,\, \hat \bn )
\left[ 1+\beta
\,\xi\,\left(
 4-n_\Omega \right)
 -\beta \,\xi_{,i}\,\left(\ln \Omega_{\rm GW}' \right)_{,i}\right]\,,
\ee
where we denote with a comma the covariant derivative, $\nabla_i \xi=\xi_{,i}$, and we  use the identity 
 $ \xi_{,i}
\,=\, \hat \bv_i- \,\xi\,\hat \bn_i$. The second term in the parenthesis of  \eqref{genexpom6a} 
is a kinematic   modulation of the rest-frame $\Omega_{\rm GW}'$; the third term is due to kinematic aberration.
% In section \ref{sec-prospects}
%we  will make  use  of an expression equivalent to \eqref{genexpom6a}.  

\smallskip

We can also directly expand eq \eqref{genexpom5a} in powers of $\beta$, and
use a 
spherical harmonic decomposition 
so  to understand  in a more transparent way the physical implication of a Doppler boost. We will carry on 
a more general analysis of these topics in section \ref{sec-prospects}.
To acquire familiarity with physical
consequences of boosting rest-frame anisotropies,
  here we present kinematic effects up to the quadrupole $\ell=2$
expanding at second order in $\beta$.
%to acquire familiarity with physical
%consequences of boosting rest-frame anisotropies. 

 %kinematic effects on intrinsic 
 %anisotropies,
% we   apply  a spherical harmonic decomposition. 
  We express the rest-frame function $ \Phi'(  \hat\bn'   )$ appearing in eq \eqref{factcon1} as
 \be
 \Phi'(  \hat\bn'   )\,=\,\sqrt{4 \pi}\,\sum_{\ell=0}^2\,\sum_{m=-\ell}^\ell\,\Phi'_{\ell m}\,Y_{\ell m} (\theta,\varphi)\,,
 \ee
where $\Phi'_{\ell m}$ are the constant coefficients of the spherical harmonic decomposition in the rest
frame ${\cal S}'$ of the SGWB. They are frequency-independent given the factorization hypothesis of \eqref{factcon1}.
 %We limit our expansion up to the quadrupole, $\ell=2$. 
 We assume   a unit monopole coefficient $\Phi_{00}'\,=\,1$, factorising it in the overall
 frequency-dependent 
 factor.  
We implement this decomposition in eq \eqref{genexpom5a},
 expanding up to order $\beta^2$. As done in the previous section, we assume  that $\hat \bv$
points towards the $e_z$ direction. We expand the GW density parameter up to the quadrupole in the basis  of  spherical harmonics, finding
\be
\Omega_{\rm GW}(f,  \hat \bn )\,=\, \sqrt{4 \pi}\,
\Omega'(f)
\,\sum_{\ell=0}^2\,\sum_{m=-\ell}^\ell\,\Phi_{\ell m}(f)\,Y_{\ell m} (\theta,\varphi)\,,
\ee
with the following non-vanishing anisotropy coefficients %{\color{red} \bf COMPLETE}
\bea\label{Phin}
\Phi_{00}&=&1+\frac{\beta}{\sqrt 3} \left(2-n_\Omega \right)\Phi_{10}'+
\frac{\beta^2}{6} \left[ 8+n_\Omega \left( n_\Omega-6\right)
+\alpha_\Omega
\right]
%\nonumber
%\\
%&&
+\frac{\sqrt{5}\,\beta^2}{12}\left[50+ (n_\Omega-21 ) n_\Omega   +\alpha_\Omega \right]\Phi_{20}'\,,
\nonumber
\\
\\
\Phi_{10}&=&
\Phi_{10}'+\frac{\beta}{\sqrt 3}  \left(4-n_\Omega\right)\,+ \frac{\sqrt{5}\,\beta}{\sqrt{12}} \left(n_\Omega-10\right)\Phi_{20}'
%\nonumber
%\\
%&&
 +\frac{\beta^2}{2}
\left(3 n_\Omega-14 \right)
\Phi_{10}'\,,
\\
\nonumber
\\
\Phi_{20}&=&\Phi_{20}'+\frac{2 \beta}{\sqrt{15}} \left(5- n_\Omega\right)\Phi_{10}'
+\frac{\beta^2}{3 \sqrt 5}\left(20-{9 n_\Omega} +{n_\Omega^2}+{\alpha_\Omega}\right)
%\,\frac{\Phi_{00}'}
%\nonumber
%\\
%&&
-\frac{\beta^2}{6}  \left[80 + \left( n_\Omega-24 \right) n_\Omega + \alpha_\Omega \right]
{\Phi_{20}'}\,.
\nonumber
\\
\label{quadgene1}
%\varphi_2
\eea
%\textcolor{red}{  For $m\neq0$ we have $\Phi_{\ell m}\,=\,\Phi'_{\ell m}$, given our specific choice of direction $\hat \bv$ (see the comment at the end of section \ref{sec-example1}). I do not understand this :-(}
  
 \noindent  
The previous expressions \footnote{ In general,  the Taylor expansions we consider always converge to the formula in eq \eqref{genexpom4} we started from, so they are  mathematically consistent. However, if the tilts of the SGWB are too large, there is the risk that the terms we neglect in the expansions are of the same order than the ones   we consider, hence our truncations can be physically misleading. In what follows, we will   consider  our formulas to be valid in regimes where  the terms we neglect in the Taylor expansions are hierarchically smaller than the contributions we include. \label{footn_discl}} have  interesting
properties:

\begin{itemize}
\item
 The  monopole  $\Phi_{00}$ in the  moving frame  is modulated by boost
 induced contributions  at order $\beta^2$ in the expansion, as well as new parts  inherited from the rest-frame dipole $\Phi'_{10}$ and quadrupole $\Phi'_{20}$.
  Notice that  there is a contribution at  order $\beta^1$, induced by the intrinsic dipole $\Phi'_{10}$, which is generally the largest
  in size given 
   our hypothesis of a  small-$\beta$ expansion. 
\item
Both the dipole $\Phi_{10}$ and quadrupole $\Phi_{20}$  receive  kinematic modulations
of their rest-frame amplitudes, as well as a kinematic aberration 
depending on the amplitude of the %remaining
  rest-frame quantities $\Phi'_{10}$ and $\Phi'_{20}$.
 Besides a kinematic modulation at  order % \textcolor{magenta}
 {$\beta^2$} to the dipole, 
 new aberration effects
 arise at order $\beta^1$, which  depend  on the size of existing rest-frame anisotropies. 
% and the anisotropies in the rest frame. 
%\item {\color{magenta} In general,  the Taylor expansions we consider always converge to the formula in eq \eqref{genexpom4} we started from, so they are  mathematically consistent. However, if the tilts of the SGWB are too large, there is the risk that the terms we neglect in the expansion are of the same order than the first two terms that we consider, hence our truncations can be physically misleading. In what follows, we will   consider  our formulas to be valid in regimes where  the terms we neglect in the Taylor expansions are hierarchically smaller than the contributions we include. }
\end{itemize}
% We notice that while the
%terms propto $\beta$ depend on the slope of the spectrum through the quantity $n_\Omega$, the
%the terms propto  $\beta^2$ depend also on $\alpha_\Omega$. 
Hence 
%already at first order in $\beta$, 
a Doppler  boost introduces kinematic aberrations that mix
different orders in a multipole expansion. For example, at order $\beta^1$, the moving frame dipole coefficient
$\Phi_{10}$ receives contributions from the rest-frame quadrupole $\Phi_{20}'$, and 
 the moving frame quadrupole coefficient
$\Phi_{20}$ receives contributions from the rest-frame dipole $\Phi_{10}'$.
We will meet again and make use of  this phenomenon in section \ref{sec-prospects}.
{ The formulas we derived in section \ref{sec-example1} and this section \ref{sec-example2}, starting
from the general result in eq \eqref{genexpom4}, go beyond what previously done in the literature, for example by considering a Taylor expansion beyond the linear order in $\beta$ (see e.g. \cite{Jenkins:2018lvb} for linear order expressions),  and including the effects of intrinsic anisotropies, not discussed in \cite{Bartolo:2022pez}.}

 The modulation and aberrations effects can be amplified in models with an enhanced slope of the GW spectrum
in certain range of frequencies,  and/or in scenarios
with intrinsic large anisotropies in the rest frame. 
 This suggests that Doppler effects can be used as a complementary probe of the SGWB frequency profile,
 as well as of its intrinsic rest-frame anisotropies.  We elaborate on this topic in what comes next.

%\subsection{Connecting with Giulia's notes}

%Here we focus up to order $\beta^1$ only. 
%Starting from \eqref{genexpom4aB}, we define 
%\bea
%\bar \Omega_{GW} (\omega)&=& \frac{1}{4 \pi}\,\int d^2 \hat \bn\,\Omega_{\rm GW}(\omega,  \hat \bn )
%\\
%&=&\Omega'(\omega)
%\left[ 1+\,M(\omega)\right]
%\\
%&=&\Omega'(\omega)
%\left[ 1+\frac{\beta}{3} \left(2-n_\Omega \right)\,\varphi_1\right]
%\eea
%Then at order $\beta$ the average of GW energy density receives a contribution from the rest-frame
%dipole $\varphi_1$. 

%We then define a quantity $\Theta(\omega, \hat n)$ that controls the relative anisotropies with respect
%to the average, up to order $\beta$, and compute it using eq \eqref{genexpom4aB}:
%\bea
%\Theta(\omega,\,\hat \bn)&\equiv&\frac{\Omega_{\rm GW}(\omega,  \hat \bn )-\bar \Omega_{GW} (\omega)}{\bar \Omega_{GW} (\omega)}
%\\
%&=& \, \hat \bn \, \hat \bv\,\left\{ \varphi_1+\beta\left[ 4-n_\Omega+\frac{\varphi_1^2}{3} (n_\Omega-2) 
%+\frac{\varphi_2}{3} (n_\Omega-10) \right] \right\}
%\nonumber
%\\&+&
 %\left( (\hat \bn \,\hat \bv)^2-\frac13   \right)
 %\left\{\varphi_2+\beta
 %\,\varphi_1\,
 %\left[ 5-n_\Omega
%+\frac{\varphi_2}{3} (n_\Omega-2) \right]\right\}
%\eea
%Anisotropies associated with $\Theta(\omega,\,\hat \bn)$ depend on  
%rest-frame stochastic anisotropies $\varphi_1$ and $\varphi_2$ of $\Omega_{GW} (\omega)$,
%as well as kinetic-induced effects propto $\beta$, that can depend or not on the rest-frame
%anisotropies. 

\section{ Kinematic anisotropies and SGWB scenarios}\label{sec-knees}

In this section we  explore cosmological and astrophysical scenarios where our previous findings can be applied.
 %start discussing applications 
%of the previous findings to different SGWB scenarios. 
 We are especially interested in theoretically identifying set-ups where
  kinematic effects can be amplified  in frequency ranges probed by GW experiments, {
  thanks to  enhancements of 
the tilts $n_\Omega$, $\alpha_\Omega$ of the SGWB spectrum at particular frequencies.}  
%
%
 %We  consider
 %representative frameworks
%apply the previous findings to SGWB scenarios
 % where Doppler effects can be relevant,  being 
  %in scenarios
  %In particular,
 %we 
 % discuss examples in which
%  where the kinematic
%anisotropies analysed in the previous section can be 
%enhanced by features
%in the frequency dependence of the GW density parameter $\Omega_{\rm GW}$.  
 %This phenomenon occurs in frequency ranges  where sources pass  from being
%effective to less-effective in producing  a  SGWB. 
We focus on  the SGWB from inflation (subsection \ref{sec-ex-infl}) and from astrophysical sources
(subsection \ref{sec-ex-astro}).
% We  %do not intend to be exhaustive, but only to 
%briefly 
%discussing representative examples
%of the phenomena we described in the previous section. 

\subsection{Primordial SGWB from the early universe}\label{sec-ex-infl}

In analogy to what happens for the CMB, we expect  also the SGWB to be
 characterized by  kinematic anisotropies due to the motion of the
solar system with respect to the cosmic rest frame  with velocity
 %  parameter 
$\beta\,=\,1.23 \times 10^{-3}$.  
%\textcolor{red}{Not sure this is correct...it is a black body spectrum and size of ansitropies is the one of vanilla cosmo GW backgrounds... While the spectrum of temperature fluctuations
%is nearly flat in the frequency range probed by the CMB -- hence spectral
%tilts do not play too an important role in characterizing kinematic anisotropies --
%for the case of a SGWB more general possibilities can occur.} In fact, 
Several
early universe models predict rich slopes in frequency for the spectrum
of  $\Omega_{\rm GW}$ which can be probed at interferometer
scales (see e.g. \cite{Bartolo:2016ami} for a study in the context of LISA). In these models, spectral tilts can become 
large enough  to compensate for  the smallness of $\beta$ in our Taylor expansion. 

Focussing on a frequency range { that can be probed} with space-based
or ground-based interferometers, one finds 
(see e.g. \cite{Maggiore:2018sht}):
\be \label{genF1}
h_0^2\,\Omega_{\rm GW}(f)\,=\,6.73\times 10^{-7}\,\,{\cal P}_T(f)\,,
\ee
which shows that  $\Omega_{\rm GW}(f)$ is  proportional to the primordial isotropic spectrum 
of tensor modes, defined as $${\cal P}_T(f)
\,=\,\frac{k^3}{2 \pi}\, \langle h_{ij}^2\rangle'\,,$$ where a prime  indicates 2-point correlators understanding
the momentum-conserving Dirac delta. For
scale-invariant and power-law primordial tensor spectra with constant slope (which can 
be relatively large and blue-tilted in models of supersolid inflation, see e.g. \cite{Cannone:2014uqa,Bartolo:2015qvr,Ricciardone:2016lym}), it is simple
to obtain the tilts of the spectrum using eq  \eqref{defno} and \eqref{defao}, and
to compute the   expressions for the Doppler anisotropies.

\medskip

 We analyze here a slightly less straightforward
example, so to explore physically well-motivated situations where the frequency-dependence of the spectrum is richer.
We consider   a primordial
SGWB sourced at second order in perturbations from scalar
fluctuations enhanced at small scales, a subject first explored in   \cite{Matarrese:1992rp,Matarrese:1993zf,Matarrese:1997ay,Noh:2004bc,Nakamura:2004rm,Ananda:2006af,Baumann:2007zm}.  These scenarios
arise frequently in models leading to primordial black hole production,
see e.g. \cite{Sasaki_2018,Carr_2016} for recent reviews.
 Such   source
 can induce a rich frequency dependence in the tensor spectrum, particularly when the width of the scalar spectrum  %${\cal P}_{\psi}$ 
  is
 small and centered at a characteristic frequency $f_\star$  (see e.g. \cite{Saito:2008jc,Saito:2009jt}). 
\begin{figure}
\centering
  \includegraphics[width = 0.5 \textwidth]{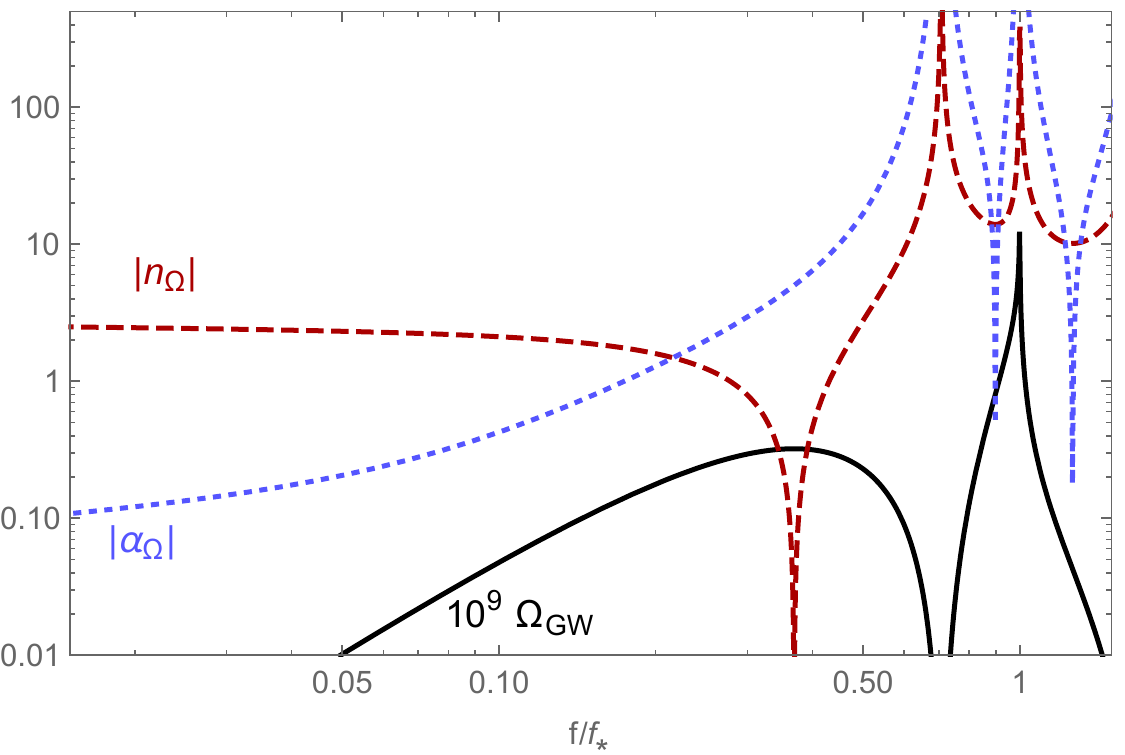}
 \caption{\small Plot of  the GW density parameter $\Omega_{\rm GW}$ obtained
 from the scalar spectrum of   \eqref{lognan1} with  width $\Delta = 0.2$. We also
 represent the absolute value of its tilts $n_\Omega$ and $\alpha_\Omega$. The
 tilts become   large at the location of features of the spectrum.}
 \label{fig:plot1}
\end{figure}
%
%But in many models  an inflationary SGWB with
 %
% Many of the inflationary models whose tensor fluctuation
%spectrum has 
%an amplitude
%  an amplitude of 
 % tensor fluctuations
  %at amplitudes
  % that can be probed
 %at interferometer scales  is sourced at second order by an enhanced scalar spectrum
 For example, we can parameterise the scalar spectrum  ${\cal P}_{\psi}$ in 
terms of log-normal Gaussian peak in frequency,    as
%  \cite{Pi:2020otn}:
\be
\label{lognan1}
{\cal P}_{\psi} (f)\,=\,\frac{{\cal A}}{\sqrt{2 \pi} \Delta }\,\exp{\left\{
-\frac{
\left[ \ln(f/f_\star) \right]^2}{2 \Delta^2}
\right\}
}\,.
\ee
Such Ansatz  leads to fully analytical formulas for $\Omega_{\rm GW}$,
as shown in \cite{Pi:2020otn}
 building
on the works of \cite{Ananda:2006af,Baumann:2007zm,Saito:2008jc,Saito:2009jt,Espinosa:2018eve,Kohri:2018awv}. In eq \eqref{lognan1}, ${\cal A}$ is the amplitude of the peak, $\Delta$ its width, and $f_\star$ a
characteristic frequency. The resulting GW spectrum has a rich and steep profile in frequency if $\Delta\ll f_\star$.
We represent the profile of the induced  $\Omega_{\rm GW}$ in Fig \ref{fig:plot1}, using the analytic results of \cite{Pi:2020otn}.
We also plot the absolute value of the parameters $n_\Omega$ and $\alpha_\Omega$, as defined in \eqref{defno},
\eqref{defao}.  A scalar spectrum with a single pronounced peak as the one of eq \eqref{lognan1} can also be obtained
 in multifield inflation, see for example \cite{Braglia:2020eai}.

\smallskip

In fact, in  the limit of thin peak $\Delta\ll f_\star$, the features of $\Omega_{\rm GW}$  can be analytically understood (see e.g. \cite{Ananda:2006af,Saito:2008jc,Saito:2009jt}). The GW spectrum  starts increasing as $f^2$ from small
towards large  frequencies. It then shows  a rapid drop in power and a zero at frequencies of order $f/f_\star\,=\,\sqrt{2/3}$.  
 A resonance~\footnote{ In realistic examples, we expect the sharp
 peak at the resonance position to be smoothed out \cite{Ananda:2006af}, so we will not consider the enhancement of kinematic effects occurring precisely at the resonance frequency  $f/f_\star\,=\,2/\sqrt{3}$. \label{foot_discl2}} then produces a
  pronounced peak, occurring at frequency  $f/f_\star\,=\,2/\sqrt{3}$.  It then definitely
drops and it vanishes at frequencies $f/f_\star\,>\,2$, since,  working at second order in perturbations, momentum conservation  does not allow to generate  tensors whose momenta are  larger than twice the scalar momentum. 

\begin{figure}[h!]
\centering
    \includegraphics[width = 0.5 \textwidth]{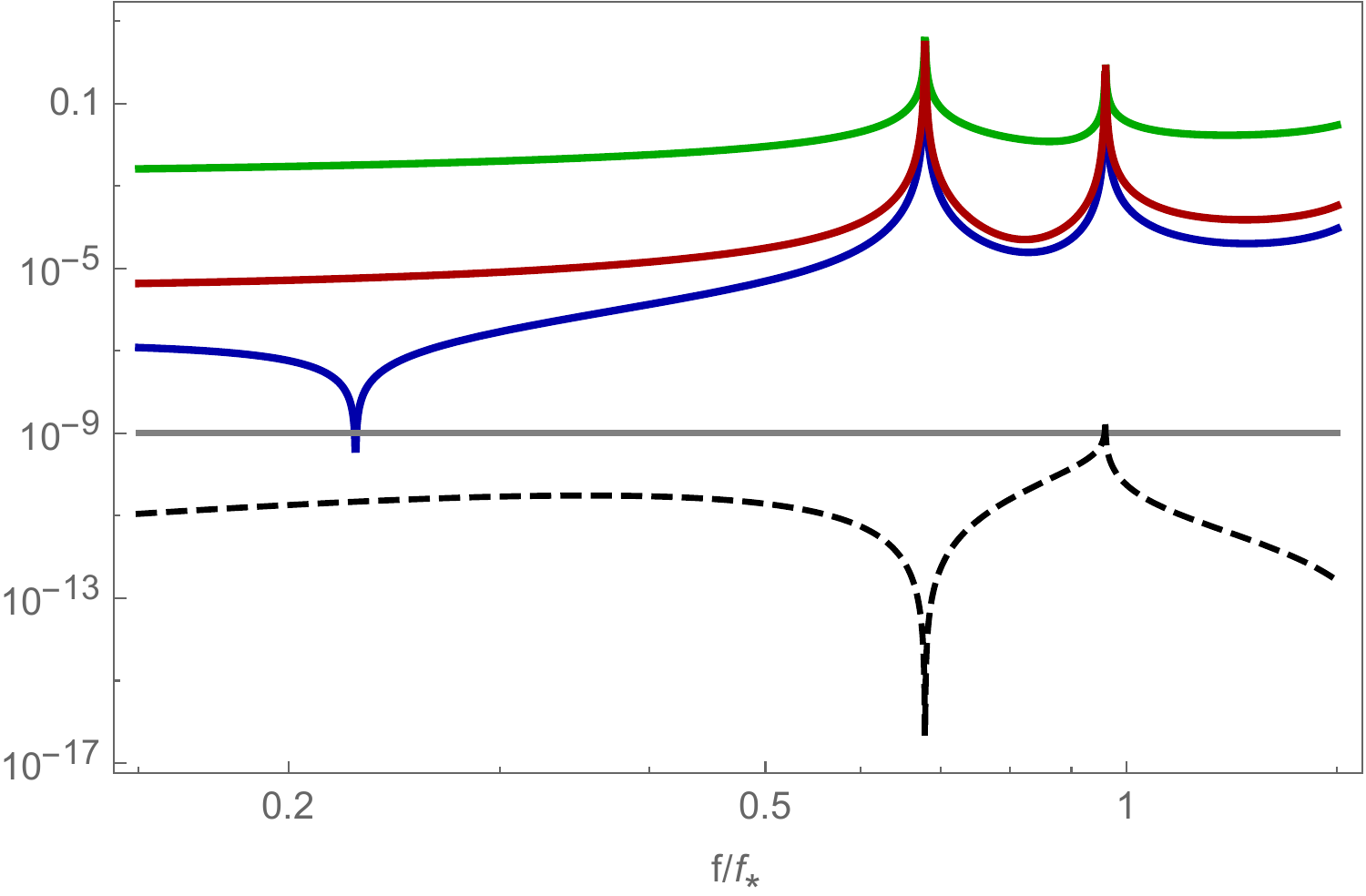}
 \caption{\small Representation of the relative contributions to eq  \eqref{genexpom4a} for the system
 GW density parameter represented in Fig \ref{fig:plot1}. We choose the CMB value for
 $\beta\,=\,1.23 \times 10^{-3}$. Blue is the monopole; Green is the dipole {divided by $\beta$}; Red  is the quadrupole {divided by $\beta^2$}.
  Below the grey line,  for reference, we include   the shape profile of the original $\Omega_{GW}$ in arbitrary units. Notice that
  the dipole contribution, starting at order $\beta^1$, is much larger than the others, and can
  be up to two orders of magnitude larger than $\beta$ within the frequency range where features in the spectrum occur. 
  }
 \label{fig:plot5}
\end{figure}

Let us assume that the primordial  anisotropies in the rest-frame ${\cal S}'$ are negligible, so to
  work in the set-up of section \ref{sec-example1}. 
  Due to the large absolute values of the tilt
parameters $n_\Omega$, $\alpha_\Omega$ we expect large  induced kinematic anisotropies 
at least in the specific frequency range $\sqrt{2/3}\,\le\,f/f_\star\,\le\,2$. 
The amplitude of the kinematic contributions to the monopole, dipole, and quadrupole
of the SGWB amplitude in the moving frame ${\cal S}$ is controlled by the functions 
$M(f)$, $D(f)$, $Q(f)$ introduced in eq \eqref{genexpom4a}.
 In figure \ref{fig:plot5}
we  plot these quantities as a function of  frequency, showing that they are indeed enhanced
in the expected frequency interval: the dipole contribution is the dominant
one since it is weighted by a single power $\beta^1$ of the expansion parameter.  { In particular, a pronounced amplification of kinematic anisotropies occurs at the position 
of the first dip of the spectrum, around  $f/f_\star\,=\,\sqrt{2/3}$.

We should now reconsider footnote \ref{footn_discl}. Given that the spectral tilts become large where the spectrum 
has features -- see Fig \ref{fig:plot1} -- we might ask whether {the expansion in powers of $\beta$ is consistent in this context. In particular we want to check whether }higher order
contributions   to the kinematic anisotropies   can turn larger than the ones we included, thus invalidating our formulas truncated at second order in a $\beta$ expansion. 
 We discuss this issue in the technical Appendix \ref{app_issuexp}, where we show that
higher order corrections in a $\beta$ expansion are hierarchically smaller, hence the results
 plotted in Fig \ref{fig:plot5} are robust.}

It would be  interesting to explore whether a detection
of kinematically induced anisotropies is possible for these scenarios, and whether it can complement
direct measurements of the slope of the spectrum (see e.g. \cite{Kuroyanagi:2018csn,Caprini:2019pxz,Flauger:2020qyi} for methods and forecasts). 
This possibility would allow one to better
characterize the spectral profile of the SGWB.  We discuss first steps towards this aim in  section \ref{sec-prospects}. We notice that while we focussed on the consequences
of a single peak in scalar fluctuations, there are more complex models with multiple peaks, steps in the inflationary potential,  or  multifield
inflationary scenarios  where even richer features occur in the spectrum as function of frequency -- see for example \cite{Cai:2019amo,Inomata:2019ivs,Fumagalli:2020nvq,Fumagalli:2021mpc}.

% \cite{Inomata:2019ivs}.  
%Moreover, \cite{Espinosa:2018eve,Kohri:2018awv,Cai:2019amo}

Until now, we assumed that the SGWB spectrum  is perfectly isotropic
in the
rest frame. On the other hand, anisotropies are expected, both of primordial
origin, or induced by propagation effects from the early universe to today:  see e.g. \cite{Alba:2015cms,Contaldi:2016koz,Bertacca:2017vod,Bartolo:2019oiq,Bartolo:2019zvb,Bartolo:2019yeu,Domcke:2020xmn,DallArmi:2020dar,Dimastrogiovanni:2021mfs}.   
%Ricciardone:2021kel,Braglia:2021fxn,}.  
 For example, sizeable
intrinsic quadrupolar anisotropies can be produced in scenarios
with large tensor non-Gaussianity, see e.g. \cite{Bartolo:2015qvr,Ricciardone:2016lym,Ricciardone:2017kre,Dimastrogiovanni:2018gkl,Dimastrogiovanni:2019bfl,Adshead:2020bji}. It would be interesting to study effects of kinematic aberration on these intrinsic anisotropies, 
given that
 they can be induced already at order  $\beta^1$ (see
 section \ref{sec-example2}) and might then be enhanced in frequency ranges where the spectrum has enhanced tilts.  
 
% Besides a complementary characterization of the slope and of the intrinsic anisotropies of the SGWB, Doppler-induced effects
 %can probe other interesting cosmological features. 
% {\bf \color{red} EXPAND ON THIS }
  Measurements of kinematic dipolar anisotropies have also been
proposed as a method to detect the chirality of a cosmological SGWB  with planar interferometers, see \cite{Seto:2006hf,Domcke:2019zls}.
Our analysis can be extended to study how parity violating effects can influence Doppler-induced
 modulations and aberrations at higher order in a multipole expansion.
  We leave the exploration of these topics  to future studies, as well as an analysis of the impact of black hole
  binaries on the detection of the primordial background \cite{Lewicki:2021kmu}. 
 
% {\color{blue}\bf GT: maybe add something}

%Hence, within the frequency
%range $\sqrt{2/3}\,\le\,f/f_\star\,\le\,2$ we expect rapid changes in slope that amplify the kinematic effects studied
%in Section \ref{sec-example1}.
%In Figure \ref{fig:plot5} we represent the relative size of the contributions $M(f)$, $D(f)$, $Q(f)$ 
%in eq \eqref{genexpom4a}, showing that the kinematic dipole can be pushed to amplitudes of order ${\cal O}(10^{-1})$ in the frequency range discussed above.  

%In the limit of very thin peak $\Delta\ll1$, the log-normal peak can be approximated as delta-function in log-space
%\be
%{\cal P}_{\psi} (k)\,=\,{\cal A}\,\delta\left[ \ln(k/k_\star) \right]
%\ee
%In the limit of small width $\Delta$, 

\subsection{Astrophysical SGWB}
\label{sec-ex-astro}
 The astrophysical stochastic gravitational-wave background (AGWB)  is generated by the superposition of signals from various resolved and unresolved astrophysical sources from the onset of stellar activity until today see e.g.  \cite{Rosado:2011kv,Regimbau:2011rp,Pitrou:2019rjz}. 
 The AGWB from binary black hole coalescence  (BH) is expected to be dominant in the Hz band  and below \cite{Dvorkin:2016okx}, and may become a source of confusion noise for some of the other types of sources. 
 
%Observations with LISA will allow one to study some aspects of resolved and unresolved stellar-mass BH binaries that are difficult to observe with ground-based interferometers.
%For example, at the mHz frequencies accessible to LISA, some of the binaries may not be fully circularized, and their residual eccentricities may provide an indication to their formation channel. In particular, binaries formed through dynamical processes in dense stellar cluster can have measurable eccentricities. These can be constrained for the subset of resolved merger, and in addition the distribution of eccentricities of the entire population may also affect the resulting AGWB.
 
 The detection of the binary neutron star (NS) coalescence  by the
LIGO/Virgo network  \cite{TheLIGOScientific:2017qsa}, and the estimated rate  $R$
of mergers in the local Universe,  which is of order $R = 920^{+2220}_{-790}\text{Gpc}^{-3}\text{yr}^{-1}$ \cite{LIGOScientific:2018mvr},  lead to the conclusion that these sources may have a comparable contribution to the AGWB relative to binary BHs \cite{LIGOScientific:2019vic}. 
 We may therefore expect that their contribution to the anisotropies of the AGWB is also  important.
 
 It is important to stress that the AGWB in the mHz and Hz band (accessible respectively by space  and ground-based interferometers) is very different in nature. In the mHz band, we are sensitive to the inspiralling phase of the evolution of binary system of (solar mass) compact objects. The duration of the inspiralling phase is long with respect to  human time scales, hence the background in the mHz band is irreducible. In the Hz band, however, we detect the very final phase of the evolution of binary systems of compact objects: mergers are well separated in the time-domain,  with almost no  overlap in time~\footnote{In other words, with an instrument of  very high sensitivity, these events are detectable individually with a catalogue approach. For this reason the background in the Hz band is not irreducible (at least as long as we focus on black hole merger contributions).}.

\begin{figure}[h!]
\centering
  \includegraphics[width = 0.59 \textwidth]{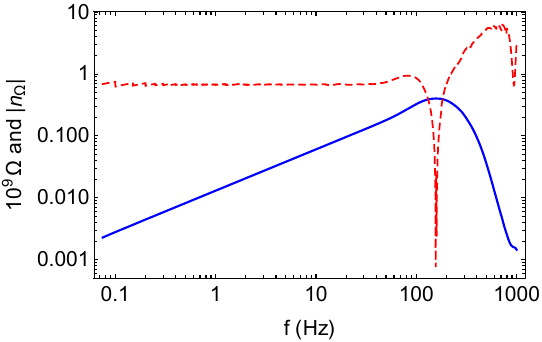}
 \caption{\small Energy density parameter as function of frequency for AGWB dominated by black hole mergers, compared with the corresponding spectral index as function of frequency (red dashed line). We choose the astrophysical model of  \cite{Cusin:2019jpv, Cusin:2019jhg} for the source population. Notice the rapid
 drop in frequencies at the coalescence stage of the BH population, leading to an increase of the tilt of the spectrum of almost an
 order of magnitude. }
 \label{fig:plot3}
\end{figure}

We represent  in Fig.\,\ref{fig:plot3}
 the  evolution of the energy density with frequency  for the astrophysical model used as a reference model in \cite{Cusin:2019jpv, Cusin:2019jhg}. In the infrared
 side of the spectrum,   the scaling with frequency follows the $\Omega_{\text{GW}}\propto f^{2/3}$ rule dictated by
 the  Einstein
   quadrupole formula {which captures the dynamics of the inspiralling phase}. The peak    in the Hz band is due to GW emission during the merger phase. While this qualitative behaviour  is universal, the width of the peak and how fast it decays depends on the details of the underlying model for mass and redshift distribution of sources.  For example, in the unrealistic scenario in which all coalescing binaries have the same mass $M$ and are located at the same distance $D$, the spectrum has a rapid drop in
   frequency at a given $f\,=\,f_{\rm drop}(M, D)$, because
   the sources of AGWB become ineffective
   at frequencies larger than $f_{\rm drop}$. In this case, the tilt of the spectrum
   can be large at frequencies around $f_{\rm drop}$, enhancing the size of kinematic effects on the AGWB
   anisotropies.
   % see Fig.\,\ref{fig:plot3}. 

 Traditionally, the energy density of the AGWB  has been modeled and parameterized under the assumption that both our universe and the distribution of sources are  homogeneous and isotropic (see e.g. Refs.~\cite{Dvorkin:2016okx,Regimbau:2011rp}). This is a rather crude approximation: GW sources are located in galaxies embedded in the cosmic web; moreover, once a GW signal is emitted, it is deflected by the presence of massive structures, such as galaxies and compact objects.  
It follows that the  energy flux from all astrophysical sources has a stochastic, anisotropic  dependence on direction. 
%From a theoretical perspective, as any background of radiation, the AGWB is fully characterized in terms of Stokes parameters, intensity and polarization, as a function of direction and frequency see e.g.~\cite{Romano:2016dpx}. 

The first prediction of the AGWB angular power spectrum was presented in ~\cite{Cusin:2018rsq, Jenkins:2018uac} following the methods   developed  in Refs.~\cite{Cusin:2017mjm, Cusin:2017fwz}. This framework is flexible and splits the cosmological large-scale structure and sub-galactic scales so that it can be applied to any source contributions and to  any frequency band. The astrophysical dependence of the angular power spectrum on the detail of the underlying astrophysical model has been studied in \cite{Jenkins:2018lvb, Cusin:2019jpv,  Cusin:2019jhg, Jenkins:2019uzp, Jenkins:2019nks} and different formal aspects of the derivation of anisotropies and their interpretation are discussed in \cite{Contaldi:2016koz,Cusin:2018avf,Bertacca:2019fnt,Pitrou:2019rjz,Alonso:2020mva}.
 %\cite{Contaldi:2016koz,  Cusin:2018avf,Bertacca:2019fnt, Pitrou:2019rjz,Alonso:2020mva}. 
 The angular power spectrum on large angular scales is characterized by the typical decay of the galaxy correlation function as a function of multipoles $\ell$, i.e. $C_{\ell}\propto (\ell+1)^{-1}$. This is not surprising as GW sources are a biased tracer of the underlying galaxy distributions and clustering is the dominant contribution to the energy density anisotropy. This implies that in the multiple expansion (\ref{Phin}), which  relates  boosted and unboosted multipoles, one gets { $\Phi'_{\ell}\sim \sqrt{(\ell-1)/\ell}\, \Phi'_{\ell-1}$}. Hence, the  intrinsic dipole and quadrupole anisotropies are typically of the same size.  In the next section we will learn that the effect of a boost is to generate an off-diagonal structure in the correlation matrix of the energy density: at a given order $\beta^n$ in an expansion in the boost velocities,  we find correlations 
 between 
 multipoles separated by $\pm n$ in their multipole indexes. 
%The relative importance of cosmological and astrophysical effects depends on the frequency band chosen, hence offering the possibility to distinguish different astrophysical processes. Due to their stochastic nature, anisotropies can be statistically characterized in terms of their angular power spectrum  and they also correlate with other cosmological observables such as weak lensing and galaxy number counts. 

 Based on the recent observations of merging black holes and neutron star binaries by the Advanced LIGO and Advanced Virgo detectors, \cite{Abbott:2021xxi}, 
 we expect  that the stochastic background from unresolved stellar-mass compact binaries may be detected within a few years of operation of the extended LIGO-Virgo network. 
 Its anisotropic component is constrained by LIGO/Virgo observations up to $\ell=4$ \cite{LIGOScientific:2019gaw},   resulting in upper limits on the amplitude of the dimensionless energy density per units of logarithmic frequency in the range $\Omega_{\rm GW}(f=25 \text{Hz},\Theta)<0.64-2.47\times 10^{-8}$ sr$^{-1}$ for a population of merging binary compact objects, where $\Theta$ denotes the angular dependence.  The updated analysis  \cite{Abbott:2021jel}  -- including also Virgo data --  improves these bounds by factors of  $2.8-3.8$.  See also \cite{Baker:2019ync}
 for a proposal of developing future large baseline interferometers for reaching higher values of multipoles $\ell$.
 %Methods to measure and map the AGWB in the LIGO and LISA frequency ranges are discussed in \cite{Allen:1996gp, Cornish:2001hg, Mitra:2007mc, Thrane:2009fp, Romano:2015uma, Romano:2016dpx, Renzini:2018vkx, TheLIGOScientific:2016xzw, 2018arXiv181108797C, Contaldi:2020rht, Alonso:2020mva}. 
 
 The study of the cross correlations with electromagnetic observables  provides complementary information and might improve the signal to noise of the anisotropic searches \cite{Cusin:2019jpv, Alonso:2020mva, Yang:2020usq}. Moreover, by  cross-correlating the GW background (which collects contribution from sources at all redshifts along the line of sight) with EM observables  at a given redshift (such as galaxy number counts), we can study  a tomographic reconstruction of the redshift distribution of sources \cite{Cusin:2018rsq, Mukherjee:2019oma,Alonso:2020mva, Cusin:2019jpv,Yang:2020usq}.

% 
%{\bf \color{red} PER GIULIA: lascio a te questa sezione, dovresti brevemente spiegare i) cosa ci si aspetta per le slopes astrofisiche ii) in che casi SGWB da inspiralling binaries possono avere 
%un drop in frequencies (essenzialmente spiegando fig \ref{fig:plot3}) iii)  magari fare una minireview delle anisotropie astrofisiche intrinseche, per spiegare como  possono essere influenzate dagli effetti cinematici. in particolare
%come la loro ampiezza scala
%  in un'espansione in multipoli (menzionando che ci aspettiamo che effetti cinematici
%mischiano i multipoli $\ell \to \ell\pm1$, come troviamo in sezioni ....). Il tutto 
% a parole, senza conti nuovi. }

\bigskip
 We close this section with a brief comment on the contribution to the AGWB
 from coalescences of  galactic sources --
 see e.g. \cite{Lamberts:2018cub,Lamberts:2019nyk} for recent detailed studies.
  In this case, the AGWB is expected to be extremely  anisotropic (see \cite{Allen:1996gp} for
  an early forecast of  detection of anisotropies), and the size of the peculiar velocity $\beta$ with respect to the solar system frame is {\it  not} necessarily  the same as the  cosmological one $\beta\,=\,1.23 \times 10^{-3}$ we considered
  above. It would be interesting to understand whether Doppler 
  effects can be used for disentangling and characterizing its properties. We leave a study
  of this topic to future work.

\section{Prospects of detection}
\label{sec-prospects}

{In the previous sections we identified  two physically relevant implications  of 
  Doppler boosting a SGWB. First, the generation of kinematic anisotropies in the moving frame ${\cal S}$ of a detector starting
from the  monopole in the rest frame ${\cal S'}$ of the emitter. Second,   the  modulation and aberration of  anisotropies in  frame  ${\cal S}$
starting from anisotropies  in  frame  ${\cal S'}$. All these effects depend on both the absolute value of the velocity $\beta$ and  on the frequency slope of the emitted GW spectrum.  In particular, non-stochastic anisotropies are a modulation of the monopole, while stochastic anisotropies are a modulation (and aberration) of intrinsic anisotropies. The latter are typically suppressed with respect to the intrinsic anisotropies, unless the spectral index of the SGWB profile is very large, hence $\beta n_{\Omega}\sim 1$. In this section  we   forecast prospects
of detecting kinematic effects using both the information contained in the stochastic and non-stochastic anisotropies~\footnote{However, in most cases -- like the vanilla example we will use for  illustration -- the stochastic part of the spectrum is not measurable, and one should focus on the non-stochastic part of the SGWB.}.}

{The fact that non-stochastic anisotropies depend on the spectral frequency shape and on the relative velocity with respect the CMB rest frame has two interesting applications.
 %
%Due to these phenomena,  
First,  a measurement of kinematic anisotropies (and in particular of the kinematic dipole)  can provide us with a complementary way to probe  the frequency-dependence
of the SGWB spectrum. This can be useful in scenarios where an extragalactic and a galactic background components are overlapped, making it difficult a spectral shape reconstruction via frequency binning \footnote{A concrete example is the galactic noise contribution to a SGWB signal in the lower  part of the LISA frequency band, see e.g. \cite{Robson:2017ayy}.}. In such a situation, measuring the  kinematic dipole helps in distinguishing the backgrounds, since the  extragalactic contributions to the SGWB  are expected to have different  intrinsic velocities
with respect to galacting ones. Second, if a spectral reconstruction is possible via binning, then, out of kinematic anisotropies, one can extract information on the value of $\beta$, in a similar (but complementary) way of what done in CMB studies. }

 \smallskip
 
{In this section we  forecast the precision associated with measurements of   boost-induced anisotropies, given a detector network. In particular, we focus on the possibility
of using the Doppler effect for measuring the tilt $n_\Omega$ of the SGWB frequency spectrum, and the absolute value of the velocity $\beta$ assuming that the spectral shape is reconstructed from the monopole. For the forecasts, we use
standard textbook methods \cite{Dodelson:2003ft} and we provide analytical expressions for the SNR of boost-induced anisotropies and for the variance associated with uncertainties
on   $n_{\Omega}$ and $\beta$.  As an illustration, we consider  a simple case study: an
  astrophysical background in the Hz (ground-based) band, measured with a detector network given by Einstein Telescope (ET) and Cosmic Explorer (CE) plus a \emph{futuristic} variation of this set-up where both the instruments have an improved strain sensitivity. }

\subsection{The multipolar decomposition}

%{\bf \color{red} SMALL INTRO NEEDED}
%We start by elaborating
%a  systematic approach for building a convenient multipolar expansion in the
%GW density parameter. 
{We now propose a multipolar expansion of the GW density parameters which allows one to study the impact of boost effects on its correlation function. 
We derive formulas which are valid for any multipole $\ell$;
 %
%do not limit ourselves to study the first few multipoles, 
however, for simplicity we include only effects up to first order $\beta^1$ in our Taylor expansion. In this sense, we go beyond what we
 did
 in section \ref{sec-trsci}, although we limit to first order in $\beta$ (see also \cite{Kosowsky:2010jm}
 for a more complete treatment suited for CMB).}
 
  %For simplicity, 
  We assume that the density parameter can be factorized into a frequency and a direction dependent component.
  % and that
 %$n_{\Omega}$ is constant.
 %the frequency dependence is  polynomial, with spectral index $n_{\Omega}$. 
%  This assumption can be easily relaxed, and our analysis generalized to  more complex frequency dependences as discussed in   section
  %\ref{sec-knees}. 
 %
 In particular, 
the density parameter $\Omega_{\rm GW}'$ in the rest frame  ${\cal S}'$ is then assumed to 
have the same form of
%be factorisable
% as in 
 eq \eqref{factcon1}:
 \be
 \label{factcon1a}
 \Omega_{\rm GW}'(f',\, \hat \bn' )\,=\,\Omega'(f') \Phi'(  \hat\bn'   )\,.
 \ee
% As mentioned above, 
% we
 %focus our attention on kinematic effects at first order $\beta^1$ in a Taylor expansion.
%characterizing anisotropies in the moving frame. 
 In order to study correlations among
 the values of $\Omega_{\rm GW}$ along different
 directions, we find  convenient to split  
% boost-induced anisotropies from kinematic effects on existing ones by separating
the  moving-frame
GW density parameter
 %in the moving frame ${\cal S}$ 
in two parts as
(see also \cite{Cusin:2016kqx,Nistane:2019yzd} for a similar analysis in the CMB context) 
\be \label{decom1gw}
\Omega_{\rm GW}(f, \bn)=\Omega_{\rm GW}^{NS}(f, \bn)+\Omega_{\rm GW}^S(f, \bn)\,.
\ee
In the definition \eqref{decom1gw} the suffix   ${ NS}$ indicates what we dub {\it non-stochastic part} of the spectrum, associated with  anisotropies
that are kinematically generated from the rest-frame monopole $\Omega'(f')$. The contribution  with suffix  $S$ is the stochastic part of the spectrum, related
with (stochastically-distributed)  intrinsic anisotropies: this part   experiences  modulation and aberration effects due to the kinematic boost.
 %as we learned in the previous sections. 

At  first order in a $\beta$ expansion,
the only non-stochastic contribution is a  dipole induced by the monopole, that reads
%Once expanding in spherical harmonics, as we learned in section \ref{sec-example1}, the non-stochastic
%part of the spectrum receives a kinematically-induced dipole starting from the monopole: expanding in spherical harmonics, the coefficient
%of the dipolar term $Y_{10}$ in the expansion of the non-stochastic part of the  spectrum is
% ($\Omega'(f) $
%is the quantity appearing in eq \eqref{factcon1a})
\be\label{NS}
\Omega_{\rm GW}^{NS}(f, \hat \bn)\,=\,\sqrt{\frac{4 \pi}{3}}\,\beta\,\left(4-n_\Omega \right)\,\Omega'(f) \textcolor{black}{Y_{10}(\hat \bn)}\,,
\ee
where $\Omega'(f) $
is the quantity appearing in eq \eqref{factcon1a}. Notice that the overall coefficient in eq \eqref{NS}
depends explicitly on $n_\Omega$. 

%The non-stochastic part of the spectrum can be expanded in terms of Legendre polynomials as
%\be
%\Omega_{GW}^{NS}(\bn)=\sum_{\ell} \beta^\ell\,a_{\ell} P_{\ell}(\bn\cdot \hat{\bv})\,.
%\ee

The stochastic part of the spectrum in the frame ${\cal S}$  can be obtained from eq
\eqref{genexpom6a}, which we rewrite  in a slightly different form that is  more convenient
for our present purposes:

\be
\label{expnSgw}
\Omega_{\rm GW}^S(f, \hat \bn)\,=\,\left(1+\beta\,(4-n_\Omega)\xi \right)\left[
\Omega'_{\rm GW}(f', \hat \bn')
-\beta \left(\nabla_a \xi \right)\nabla^a \Omega'_{\rm GW}(f', \hat \bn') \right]\,. 
\ee
This expression makes  manifest the (overall) effects of modulation, and the effects of aberration
in the covariant derivatives of $\Omega'_{\rm GW}$. 
%As stated above,
%we are interested here in the effects at first order in an expansion in $\beta$.  

With these tools we  can compute correlations among stochastic anisotropies. It is convenient to expand the   stochastic
contributions to the spectrum  in spherical harmonics
\be
\Omega_{\rm GW}^{S}(f, \hat \bn)=\sum_{\ell m} \Omega_{\ell m}(f)\, Y_{\ell m}(\hat \bn)\,, \quad\Omega_{\rm GW}'(f, \bn)=\sum_{\ell m} \Omega'_{\ell m}(f)\,Y_{\ell m}(\hat \bn) \,.
\ee
Moreover, we also expand in dipolar harmonics the parameter $\xi( \hat \bn)$ of eq \eqref{defmu}
and also
\be
\beta\,\xi( \hat \bn)=\sum_{m} Y_{1 m}(\hat \bn) \beta_{1 m}\,.
\ee

\begin{figure}[htt!]
		\begin{center}
			%\captionsetup{justification=centering}
			\includegraphics[width=0.475\columnwidth]{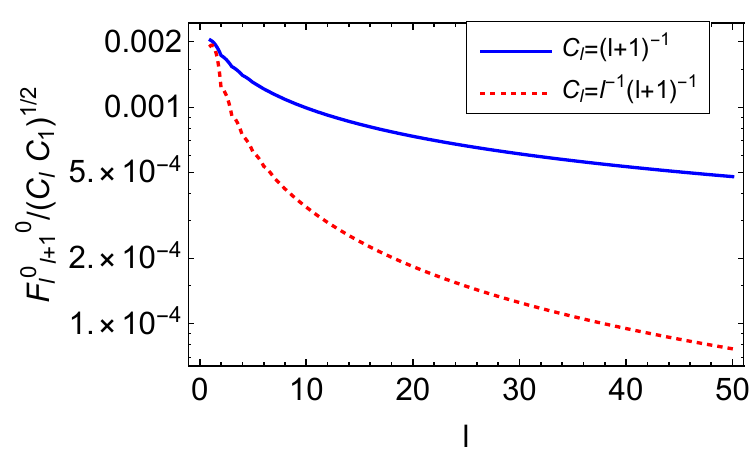}\qquad 	\includegraphics[width=0.475\columnwidth]{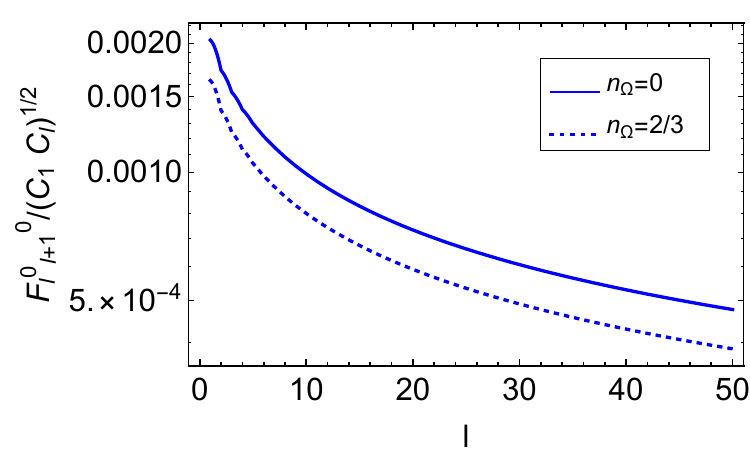}
			\caption{\small \label{FF} Left: The 
			first off diagonal term of the correlation function  \eqref{F}, as function of $\ell$. We choose $n_\Omega=0$  and we test two different scalings of the intrinsic spectrum in eq.\,(\ref{F}). Right: Same figure for two different values of $n_\Omega$ and for the astrophysically-motivated scaling of the left panel. In both panels the normalisation has been chosen for future convenience. }
\end{center}
	\end{figure}

Plugging these expansions eq \eqref{expnSgw},
after standard manipulations (see also  appendix \ref{lm}), one obtains at linear order in $\beta$ 
the following relation among coefficients,  %\cite{Cusin:2016kqx} 
\be\label{def1olm}
 \Omega_{\ell m}= \Omega_{\ell m}'+\sum_{\ell_1 m_1 m_2}\left[3-n_\Omega-\frac{\ell_1}{2}(\ell_1+1)+\frac{\ell}{2}(\ell+1)\right]\Omega_{\ell_1 m_1}'\beta_{1m_2}\mathcal{W}_{\ell \ell_1 1}^{m m_1m_2}\,. 
 \ee
In writing this formula  we introduced the Wigner-like symbol (see appendix \ref{lm})
\be
\mathcal{W}_{\ell_1\,\,\ell_2\,\,\ell_3}^{m_1\,\,m_2\,\,m_3}=\int d^2n\,Y^*_{\ell_1 m_1}Y_{\ell_2 m_2}Y_{\ell_3 m_3}\,.
\ee
Since a boost violates statistical isotropy, the  correlation functions among
the $ \Omega_{\ell m}$ of eq \eqref{def1olm} can have
%in full generality has
 a non-diagonal structure. We denote such correlations by a four-index quantity $F_{\ell m}^{\ell' m'}$ as
\be\label{Ftot}
F_{\ell m}^{\ell' m'}\equiv \langle \Omega_{\ell m} \Omega_{\ell' m'}\rangle\,.
\ee
We conveniently split (\ref{Ftot}) into a statistically isotropic contribution, and a boost-induced contribution  as
%by the  as 
 \be
 F_{\ell m}^{\ell' m'} =\delta_{\ell \ell'}\delta_{m m'}C_{\ell}+ ( F_{\ell m}^{\ell' m'})^{\beta}\,.
\ee
 The boost-induced part --  which violates statistical isotropy --  is given by
%\be
%( F_{\ell m}^{\ell' m'})^{\beta}=(3+\alpha_{\ell'\ell})\,\beta_{1 m-m'} \,\mathcal{W}_{\ell\,\,\ell'\,\,1}^{m\,\,m'\,\,m-m'}(C_{\ell}-C_{\ell'})\,,\label{v1}\qquad \alpha_{\ell'\ell}\equiv \frac{\ell'-\ell}{2}(\ell'+\ell+1)\,, 
%\ee
\be\label{F}
( F_{\ell m}^{\ell' m'})^{\beta}=\beta_{1\,\, (m-m')}\left[\left(3-n_\Omega\right)(C_{\ell}+C_{\ell'})+\alpha_{\ell'\ell}(C_{\ell}-C_{\ell'})\right] \,\mathcal{W}_{\ell\,\,\,\,\,\ell'\,\,\,\,\,\,1}^{m\,\,m'\,\,m-m'}\,, 
\ee
where
\be
\alpha_{\ell'\ell}\equiv \frac{\ell'}{2}(\ell'+1)-\frac{\ell}{2}(\ell+1)\,. 
\ee
We notice that the result depends on the spectral tilt $n_\Omega$: all the diagonal terms (i.e. $\ell=\ell'\,, m=m'$) of the correlation matrices $( F_{\ell m}^{\ell' m'})^{\beta}$ are vanishing. Off-diagonal correlators are non-vanishing only for $\ell'=\ell\pm1$, i.e. we have only correlation among $\ell\leftrightarrow \ell\pm 1$ multipoles.
This is the consequence of the effect explained after eq \eqref{quadgene1}:
 implementing a spherical-harmonic expansion of the density parameter, 
 kinematic aberrations introduce 
contaminations between different multipoles. 
 This result can be extended beyond linear order in $\beta$: at a generic order $n$th in the $\beta$-perturbation expansion in $\beta$ (i.e. order $\beta^n$), only off-diagonal elements separated at most by $n$ in multipole index are turned on. We plot the correlation  \eqref{F}
%The symbol $\mathcal{W}_{...}$ is  a combination of Wigner symbols, see appendix. 
    in Fig.\,\ref{FF} for two simple case studies.

\subsection{Fisher forecasts}\label{results}

We now develop Fisher forecasts for measuring 
 the spectral tilt
$n_\Omega$, exploiting  the properties of boost-induced
anisotropies.
 %induced by the relative motion among observers.
% \textcolor{magenta}{We now  $n_\Omega$ withthe kinematic Doppler effects.}
 We assume that the size of  $n_\Omega$ is
  not too large, so that
  %such that 
  the dipole  amplitude  dominates over the quadrupole 
  in a  perturbative expansion of the non-stochastic term 
(\ref{NS}). Hence we consider only  terms linear in boost velocity: this approximation is well justified as long as 
$\beta\,n_\Omega\ll 1$. 
%focus on dipole (effects linear in $\beta$ and on $\gamma$ and check which approximations are valid for $\gamma\sim 10^2$  }
The likelihood for the $\Omega_{\ell m}$ is assumed to be  of the standard multivariate Gaussian form:
%\footnote{This is the likelihood of  a model  given experimental data (that we will assume to be our model fiducial, plus an instrumental noise parametrisation)}:
\be
\ln \mathcal{L}=-\frac{1}{2}\left[\sum_{\ell m}\sum_{\ell' m'} \left(\Omega_{\ell m}^* - \Omega_{\ell m}^{NS*}\right) \left(F^{-1}\right)_{\ell  m}^{\ell' m'} \left(\Omega_{\ell' m'} -\Omega_{\ell' m'}^{NS} \right) +\ln \det\left(F\right)_{\ell m}^{\ell' m'}\right]+\text{const}\,. 
\ee
We stress that the non-stochastic anisotropies $\Omega_{\ell' m'}^{NS}$
induced by the monopole  are not inherently random. For this reason we treat them as a mean value. The covariance matrix is given by the two-point correlation function (\ref{F}). 
To simplify our notation, we introduce the following matrix form
\be
F_{\mu\nu}\equiv F_{\ell m}^{\ell' m'}\,,
\ee
where the first index corresponds to $\mu\equiv (\ell, m)$, while the second to $\nu\equiv (\ell', m')$. 
Calling $\delta F$  the specific contribution of  boost-induced anisotropies, 
we  write~\footnote{With slight abuse of notation,
we denote $C_{\mu}\,\equiv\,C_\ell$, although  this quantity does not depend on the index $m$.} 
%make use the following notation
\be\label{deltaF}
F_{\mu\nu}=\delta_{\mu\nu}C_{\mu}+\delta F_{\mu\nu}\,,
\ee
%where the $\delta F$ denotes the boost-induced anisotropies. In what follows we will also use
%the notation
%and
% for future convenience we also introduce 

The theoretical covariance matrix $F^{-1}_{\mu\nu}$ generally
 depends  on cosmological and astrophysical parameters $\lambda_{A}$.  The uncertainty associated
 with  these parameters is given by the Fisher matrix 
\be
\mathcal{F}_{A B}=\Big< -\frac{\partial^2 \ln \mathcal{L}}{\partial \lambda_{A}\partial \lambda_{B}}\Big>\,,
\ee
which can be written more explicitly as (we use the shortcut notation $\partial_A\equiv\partial/\partial \lambda_A$)
\be\label{Fisher}
\mathcal{F}_{AB}=\frac{1}{2} \sum_{\mu\nu\alpha\sigma} \left[(F^{-1})_{\mu\nu}\partial_{A} F_{\nu\sigma} (F^{-1})_{\sigma \alpha} \partial_{B} F_{\alpha \mu}\right]
+ \textcolor{black}{\sum_{\mu\nu} \partial_{A} \Omega_{\mu}^{NS*}  (F^{-1})_{\mu\nu} \partial_{B} \Omega_{\nu}^{NS}} \,.
\ee
%where we used the shortcut notation $\partial_A\equiv\partial/\partial \lambda_A$. 
%In full generality, the theoretical correlation function can be split into a diagonal contribution from primary anisotropy (with instrumental noise in) and an off-diagonal one from boost. 
We start considering  a perfect experiment with no instrumental noise,
and we  assume that the boost velocity is the same as the one measured  by CMB experiments. We then discuss in a second step how the  instrumental noise can be effectively included in the forecasts.

\subsubsection{Uncertainty on the reconstruction of $n_\Omega$}

%We make forecasts for measuring the tilt $n_\Omega$, controlling
%
%\subsection{Constraining $\gamma$}
We now forecast the precision associated with the  measurement of the spectral index parameter $n_\Omega$ controlling 
 %governing
  the amplitude of the boost-induced anisotropies. We assume
  that  $n_\Omega$  is constant and it is the only free parameter to measure: a more realistic analysis would vary also other model parameters, but our specific goal in
  this context is to investigate whether the tilt $n_\Omega$ is detectable in the most favourable setting. 

  %can be constrained. 
%\be\label{off}
%F_{\mu\nu}=C_{\mu} \delta_{\mu\nu}+A_L\delta C_{\mu\nu}+A_{\beta} \delta D_{\mu\nu}\,,
%\ee
In this case the Fisher matrix has only one element 
\be\label{defsigma}
\mathcal{F}_{n_\Omega n_\Omega}=\sigma_{n_\Omega}^{-2}
\,=\,\sum_\nu
\,{\frac{\partial \Omega_{\nu}^{NS*} }{\partial n_{\Omega}} (F^{-1})_{\nu}^{\nu}\frac{\partial \Omega_{\nu}^{NS} }{\partial n_{\Omega}}\delta_{\nu\, (1m)}}\,
+\frac{1}{2}\sum_{\mu}T_{\mu}\,.
\ee
The first  term in this formula is associated with    the non stochastic contributions given by eq. (\ref{NS}). We only keep the dipole, being the only contribution linear in $\beta$, and this explains the
Kronecker symbol $\delta_{\nu\, (1m)}$.  
The second term of eq \eqref{defsigma} is given by 
\be\label{Tmu}
T_{\mu}\equiv \sum_{\nu\sigma\rho}\delta C_{\mu\nu}(F^{-1})_{\nu\sigma}\delta C_{\sigma\rho}(F^{-1})_{\rho\mu}\,,
\ee
where with $\delta C_{\mu\nu}$ we denote the part of the correlation function (\ref{F}) proportional to $n_\Omega$:
% see eq.\,(\ref{deltaC}):
\be\label{deltaC}
\delta C_{\mu\nu}\equiv  \frac{\partial\,\delta F_{\mu\nu}}{\partial\,n_\Omega}\,.
\ee
Notice that the non-stochastic anisotropies $\Omega_\mu^{NS}$ given in eq.  (\ref{NS}) {\it do} contribute to this formula,  as they do depend on $n_\Omega$. Since %in Eq.\ (\ref{totalC}), 
 we expect the off-diagonal component to be  suppressed relative to the diagonal one,  we 
 write
\begin{align}
F_{\mu\nu}&=\sqrt{F_{\mu\mu}}\sqrt{F_{\nu\nu}}\left(\delta_{\mu\nu}+\frac{\delta F_{\mu\nu}}{\sqrt{F_{\mu\mu}}\sqrt{F_{\nu\nu}}}\right)\nn\\
&=\sqrt{C_{\mu}}\sqrt{C_{\nu}}\left(\delta_{\mu\nu}+\epsilon_{\mu\nu}\right)\,,
\end{align}
using the relation  $F_{\mu\mu}=C_{\mu}$. For the inverse of this quantity  we find % \textcolor{magenta}{Make sure that in our realistic case, this is invertible}
\be\label{inverse}
(F^{-1})_{\mu\nu}\simeq \frac{\delta_{\mu\nu}}{C_{\mu}}-\frac{\delta F_{\mu\nu}}{C_{\mu}C_{\nu}}\,,
\ee
which gives for (\ref{Tmu}) 
\be
T_{\mu}\approx  \frac{1}{C_{\mu}}\sum_{\nu}\frac{\delta C_{\mu\nu}\delta C_{\nu\mu}}{C_{\nu}}+\dots\,,
\ee
where terms of order $\sim(\delta C)^2\delta F/C^3$  are neglected.   

We now compute the contribution from the $\mu$-dependent term, 
to understand how the different terms in the Fisher matrix contribute to {the precision with which $n_{\Omega}$ can be reconstructed, i.e. $\sigma_{n_\Omega}^{-2}$ defined in (\ref{defsigma})}. Going back to the usual notation $(\mu)=(\ell, m)$
%\be
%(S/N)_{\mu}\equiv \sqrt{\frac{T_{\mu}}{2}}=\frac{1}{\sqrt{C_{\mu}}}\left[\sum_{\nu}\frac{(\delta C)_{\mu\nu}(\delta C)_{\nu\mu}}{ 2 C_{\nu}}\right]^{1/2}\,.
%\ee
%In the usual notation and taking the square, for future convenience
\be \label{Sigma2}
\left(\sigma^{-2}_{n_\Omega}\right)_{\ell m}=\frac{1}{C_{\ell}}\left[\sum_{\ell' m'}\frac{(\delta C)_{\ell m}^{\ell' m'}(\delta C)_{\ell' m'}^{\ell \,m}}{ 2 C_{\ell'}}\right]+\textcolor{black}{\delta_{\ell 1} \frac{1}{C_{1}}\left(\partial_{n_\Omega}\Omega_{1 m}^{NS*} \right)
\left(\partial_{n_\Omega}\Omega_{1 m}^{NS}
\right)
}+\mathcal{O}(\beta^3)\,.
\ee
This expression is  determined in terms of the multipoles in eq.\,(\ref{NS}),  the correlation matrix in eq.\,(\ref{F}),  and the non-stochastic dipole (\ref{NS}).
%The role of the non-stochastic term will be  relevant for setting the error bars on the determination of $n_\Omega$. 

%(Recall again that to simplify the notation we have  denoted as $\delta C$ the part of the correlation function proportional to $n_\Omega$, see eq.\,(\ref{deltaC})). 

Without loss of generality, we can choose a system of coordinates with azimuth aligned with $e_z$, where $e_z$ denotes the direction of the boost velocity.\footnote{To generalize our analysis, we can consider a rotated coordinate frame.  The rotation is described by a SO(3) matrix $R_1$ characterized by its Euler angles $(\varphi_1,\theta_1,0)$.  In the new coordinate frame  the  boost direction   is  described  by  the  unit  vector $n_1=R_1e_z$.   A  direction described by a unit vector $n$ in the old reference frame, is rotated to $R^{-1}n$ in the new one. The change to the rotated coordinate system does not change the results, however, and for the Fisher matrix analysis we will continue to use the preferred reference frame in which only the $m=0$ component of the boost potential is non-zero.\label{ftn-rot}} In such a reference frame $m=m'=0$. Moreover, 
the sum in (\ref{Sigma2}) can be further simplified recalling that only multipoles with $(\ell-\ell') =\pm 1$ are correlated. We then obtain, using eq.\,(\ref{NS}), 
%\be \label{SNR2bis}
%(S/N)^2_{\ell}=\frac{1}{C_{\ell}}\left[\frac{(\delta C)_{\ell 0}^{\ell+1 0}(\delta C)_{\ell+1 0}^{\ell 0}}{ 2 C_{\ell+1}}+\frac{(\delta C)_{\ell 0}^{\ell-1 0}(\delta C)_{\ell-1 0}^{\ell 0}}{ 2 C_{\ell-1}}\right]+\textcolor{black}{\frac{1}{C_{1}}\frac{3}{4\pi}\beta^2}\,,
%\ee
\begin{align} \label{SNR2bis}
&(\sigma^{-2}_{n_\Omega})_{\ell}=\frac{\beta^2}{C_{\ell}}\left[\frac{ (C_{\ell}+C_{\ell+1})^2}{ 2 C_{\ell+1}}(\mathcal{W}_{\ell\,\,\ell+1\,\,1}^{0\,0\,\,\,\,\,\,\,\,\,0})^2+\frac{ (C_{\ell}+C_{\ell-1})^2}{ 2 C_{\ell-1}}  (\mathcal{W}_{\ell\,\,\ell-1\,\,1}^{0\,0\,\,\,\,\,\,\,\,\,0})^2\right]\,,\qquad \ell>1
\\  \label{SNR2bis2}
&(\sigma^{-2}_{n_\Omega})_{1}=\frac{\beta^2}{C_{1}}\left[\frac{ (C_{1}+C_{2})^2}{ 2 C_{2}}(\mathcal{W}_{1\,\,2\,\,1}^{0\,0\,\,0})^2\right]+\textcolor{black}{\frac{4\pi}{3C_{1}}\beta^2 \left(\Omega'(f)\right)^2}\,.
\end{align}
This result confirms that,  %we see that 
at leading order in $\beta$, non-stochastic anisotropies contribute only to the dipole
through the last term in eq \eqref{SNR2bis2}. 
Hence, when considering an instrument with  angular resolution  $\ell$, the
corresponding
 uncertainty on the spectral tilt  $n_\Omega$ is given by 
\be\sigma_{n_\Omega}=\left[\sum_{\ell'=1}^{\ell} \left(\sigma_{n_\Omega}^{-2}\right)_{\ell'}\right]^{-1/2}\,.
\ee
\textcolor{black}{We stress that until now we  assumed that instrumental noise is negligible, and we
only considered effects of cosmic variance. The contribution of instrumental noise can be taken into account replacing $C_{\ell}\rightarrow C_{\ell}+N_{\ell}$  in the denominator of (\ref{Sigma2}), where $N_{\ell}$ is an estimate of instrumental noise per multipole for a given detector network, see e.g. \cite{Alonso:2020rar}.}

\subsubsection{Constraining the velocity $\beta$}

{We now assume that from the study of the monopole, we have a good reconstruction of the spectral index $n_{\Omega}$ in a given frequency band. We ask ourself the questions: can kinematic anisotropies be used to set constraints on the velocity of our relative motion with respect to the \emph{emission} rest frame? }

We assume that  $\beta$  is constant and it is the only free parameter to measure: a more realistic analysis would vary also other model parameters, but our specific goal in
  this context is to investigate whether $\beta$ can be constrained in the most favourable setting.  
  
 By repeating steps totally analogous to what done in the previous section, one finds that the variance associate to $\beta$ is given by 
 \begin{align} \label{beta}
(\sigma^{-2}_{\beta})_{\ell}&=\left[(3-n_{\Omega}) (C_{\ell}+C_{\ell+1})+(\ell+1) (C_{\ell}-C_{\ell+1})\right]^2\frac{(\mathcal{W}_{\ell\,\,\ell+1\,\,1}^{0\,0\,\,\,\,\,\,\,\,\,0})^2}{2 C_{\ell+1}C_{\ell}}+\nn\\
&\qquad+\left[(3-n_{\Omega}) (C_{\ell}+C_{\ell-1})-\ell(C_{\ell}-C_{\ell-1})\right]^2\frac{(\mathcal{W}_{\ell\,\,\ell-1\,\,1}^{0\,0\,\,\,\,\,\,\,\,\,0})^2}{2 C_{\ell-1}C_{\ell}}\,,\qquad \ell>1\,,\\
(\sigma^{-2}_{n_\Omega})_{1}&=\left[(3-n_{\Omega}) (C_{1}+C_{2})+2 (C_{1}-C_{2})\right]^2\frac{(\mathcal{W}_{1\,\,2\,\,1}^{0\,0\,\,0})^2}{2 C_{2}C_{1}}+\frac{4\pi}{3 C_1}(4-n_{\Omega})^2\Omega'(f)^2\,.
\end{align}
The contribution of instrumental noise can be taken into account replacing $C_{\ell}\rightarrow C_{\ell}+N_{\ell}$  in the denominator of (\ref{beta}), where $N_{\ell}$ is an estimate of instrumental noise per multipole for a given detector network, see e.g. \cite{Alonso:2020rar}.

  \subsubsection{Signal-to-noise ratio of boost-induced anisotropies}

We  now compute the cumulative SNR associated with  measurements of  boost-induced anisotropies. To do so, we introduce a book-keeping parameter $A$ controlling the amplitude of  boost induced anisotropies, and send $\beta\rightarrow \beta A$. Then by definition 
\be
\mathcal{F}_{AA}=\sigma_{A}^{-2}=\left(S/N\right)^2=\frac{1}{2}\sum_{\mu}P_{\mu}+\textcolor{black}{\Omega_{\nu}^{NS*}  (F^{-1})_{\nu}^{\nu}\Omega_{\nu}^{NS}\delta_{\nu\, (1m)}}\,.
\ee
where the index $A$ corresponds to the aforementioned book-keeping parameter relative to boost-induced quantities. 
We denote
\be\label{Pmu}
P_{\mu}\equiv \sum_{\nu\sigma\rho}\delta F_{\mu\nu}(F^{-1})_{\nu\sigma}\delta F_{\sigma\rho}(F^{-1})_{\rho\mu}\,,
\ee
\textcolor{black}{where $\delta F$ is defined in (\ref{deltaF})}. Eq.\,(\ref{Pmu}) can be approximated as 
\be
P_{\mu}\approx  \frac{1}{C_{\mu}}\sum_{\nu}\frac{\delta F_{\mu\nu}\delta F_{\nu\mu}}{C_{\nu}}+\dots\,,
\ee
where we use (\ref{inverse}) and as above we neglect terms of order $\sim(\delta F)^3/C^3$  and higher. Then
\be \label{SNR2}
\left(S/N\right)^2_{\ell m}=\frac{1}{C_{\ell}}\left[\sum_{\ell' m'}\frac{(\delta F)_{\ell m}^{\ell' m'}(\delta F)_{\ell' m'}^{\ell m}}{ 2 C_{\ell'}}\right]+\textcolor{black}{\delta_{\ell 1} \frac{1}{C_{1}}\left(\partial_{A}\Omega_{1 m}^{NS*}\right)\left(\partial_{A}\Omega_{1 m}^{NS}\right)}+\mathcal{O}(\beta^3)\,.
\ee

In our reference frame, $m=m'=0$. Recalling that only multipoles  $(\ell-\ell')\,=\,\pm 1$ correlate, (\ref{SNR2}) can be written as 
\begin{align} \label{SNR2bis}
&(S/N)^2_{\ell}=\frac{\beta^2}{2 C_{\ell}  C_{\ell+1}}\left[(3-n_\Omega)^2 (C_{\ell}+C_{\ell+1})^2+(\ell+1)^2 (C_{\ell}-C_{\ell+1})^2\right](\mathcal{W}_{\ell\,\,\ell+1\,\,1}^{0\,0\,\,\,\,\,\,\,\,\,0})^2\nn\\
&\qquad \qquad +\frac{\beta^2}{2 C_{\ell}  C_{\ell-1}}\left[(3-n_\Omega)^2 (C_{\ell}+C_{\ell-1})^2+\ell^2 (C_{\ell}-C_{\ell-1})^2\right](\mathcal{W}_{\ell\,\,\ell-1\,\,1}^{0\,0\,\,\,\,\,\,\,\,\,0})^2\nn\\
&(S/N)^2_{1}=\frac{\beta^2}{2 C_{1} C_2}\left[(3-n_\Omega)^2 (C_{1}+C_{2})^2+4(C_{1}-C_{2})^2 \right](\mathcal{W}_{1\,\,2\,\,1}^{0\,\,0\,\,0})^2+\frac{4\pi}{3C_{1}}\beta^2(4-n_\Omega)^2 \left(\Omega'(f)\right)^2\,,
\end{align}
where we used eq.\,(\ref{NS}). 
The cumulative SNR is given by
\be
\left(\frac{S}{N}\right)^{\text{Cum}}_{\ell}=\sqrt{\sum_{\ell'=1}^{\ell}\left(\frac{S}{N}\right)^2_{\ell'}}\,.
\ee
\textcolor{black}{We stress that up to now we assume to work in a  cosmic-variance limited regime. However, as  mentioned above, a contribution of instrumental noise can be included in (\ref{SNR2}) replacing the denominator with $C_{\ell}\rightarrow C_{\ell}+N_{\ell}$, where $N_{\ell}$ is an estimate of instrumental noise per multipole for a given detector network, see e.g. \cite{Alonso:2020rar}.}

%{\bf \color{red} qui magari scrivi qualcosa sulle limitazioni e opportunita' future dell'esempio discusso:}

\subsection{Illustration of the forecasting method: a  case-study}

As a practical {illustration of the method}, we
apply our general results  to a specific case study. We consider an  astrophysical extra-galactic  background of solar mass compact binaries with power spectral density with constant slope and $n_\Omega=2/3$. As explained in section \ref{sec-ex-astro}, for this case we expect the angular power spectrum to scale as the galaxy correlation function %on large scales
   $\sim 1/(1+\ell)$, and anisotropies to be suppressed with respect to the isotropic component of a typical factor $\sim (1-5)\times 10^{-2}$ depending on the astrophysical model. 
The astrophysical dependence of the angular power spectrum on the detail of the underlying astrophysical model has been studied in \cite{Cusin:2018rsq, Cusin:2018avf, Jenkins:2018lvb, Jenkins:2018uac, Cusin:2019jpv, Cusin:2019jhg} and different formal aspects of the derivation of anisotropies and their interpretation are discussed in 
 \cite{Contaldi:2016koz, Cusin:2017fwz,  Cusin:2017mjm, Cusin:2018avf, Pitrou:2019rjz, Alonso:2020mva}. %Alonso:2020mva}. 
%add Bertacca:2019fnt,

We  consider the most optimistic scenario in which the amplitude of the monopole is of the order of present upper bounds in the Hz band, $\bar{\Omega}_{\rm GW}(f=25 \text{Hz})\sim 3.4\times 10^{-9}$ \cite{Abbott:2021xxi} and the angular power spectrum is suppressed with respect to monopole by a factor $10^{-3}/(\ell+1)$. %\textcolor{magenta}{I propose to discuss more extensively current bounds in the introduction} 
For a given detector network, the instrumental noise curve per multipole, $N_{\ell}$, can be obtained using results of \cite{Alonso:2020rar} and the publicly available code \emph{schNell}\footnote{\url{https://github.com/damonge/schNell}}.

		\begin{figure}[htt!]
		\begin{center}
			%\captionsetup{justification=centering}
			\includegraphics[width=0.49\columnwidth]{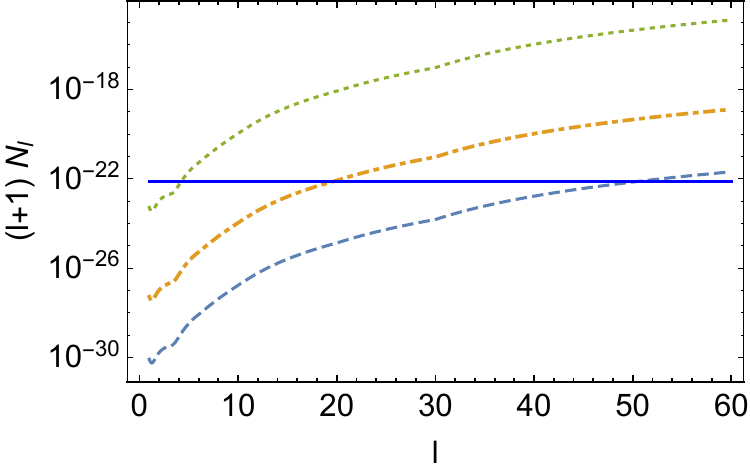}
			\caption{\small \label{Nl} Instrumental noise per multipole for CE+ET and for two futuristic scenarios where the network detectors    improve their  strain sensitivity of a factor 10 and of a factor 50 respectively. We choose an integration time $T=1$ year and $f=63$ Hz. The blue line is the expected  amplitude of a signal from a population of binary systems of stellar mass black holes, whose monopole in on the edge of being detected.  The signal is time-independent  while the noise decreases linearly with observation time \cite{Alonso:2020rar}. The
			size of the non-stochastic dipole is  $\sim 10^{-21}$. }
		\end{center}
	\end{figure}
		% In this work, 
		 For illustrative purposes we consider: 

\begin{itemize}
\item[1)] a  network made of Cosmic Explorer \cite{Reitze:2019iox} plus Einstein Telescope \cite{hild2008pushing},  together with two futuristic scenarios where
 \item[2)] both these instruments have an improvement in strain sensitivity of a factor 10 
and 
 \item[3)] of a factor 50 with respect to their  nominal values. 
 \end{itemize}
 
The instrumental noise per multipole in each of these three scenarios is plotted in Fig.\,\ref{Nl} for a pivot frequency of 63 Hz, compared with the typical amplitude of the signal associated to an extra-galactic background at this frequency (clustering component).

\medskip
 
\begin{figure}[htt!]
		\begin{center}
			%\captionsetup{justification=centering}
			\includegraphics[width=0.44\columnwidth]{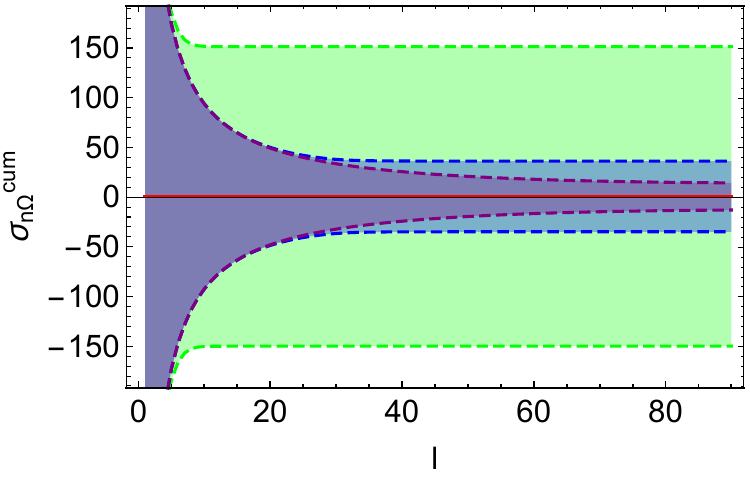}\qquad 	\includegraphics[width=0.44\columnwidth]{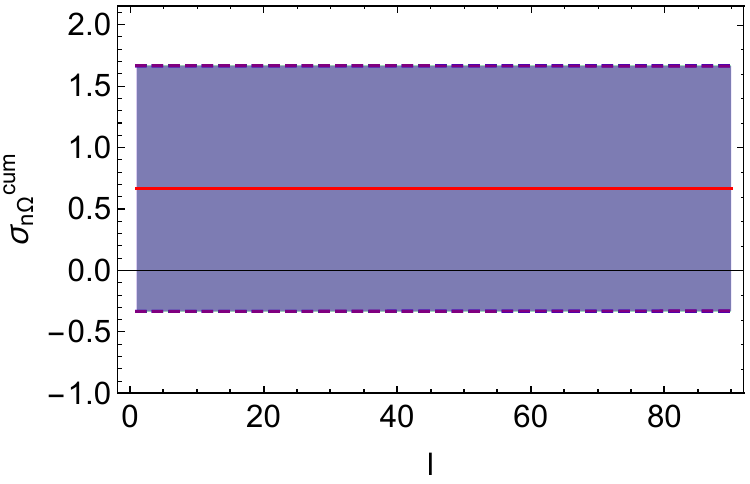}
			\caption{\small \label{SigmaCum} Variance on $n_\Omega$ as function of angular resolution. The red line represents the value of the spectral index $n_\Omega=2/3$. The shaded green, blue and purple areas correspond to the $1 \sigma$ region for the realistic scenario CE+ET and for the two futuristic scenarios described in the text (improvement in strain sensitivity of a factor 10 and 50 respectively). In the left panel we consider only the contribution of stochastic anisotropies, in the right one we also add the contribution from the non-stochastic dipole
			that plays an important role in controlling the error bars. The curves are indistinguishable up to very high multipoles, where the contribution of the stochastic component starts to be visible. We assume an integration time $T=10$ years.}
		\end{center}
	\end{figure}

In Fig.\,\ref{SigmaCum} we represent the expected (cumulative)  precision  for constraining the spectral density $n_\Omega$ in each one of the scenarios under study, as a function of multipole $\ell$.  We show separately what can be achieved using stochastic anisotropies only (left panel), and adding the non-stochastic dipole (right panel).  In the right panel, we notice  that the non-stochastic dipole sets the size of the error bar and stochastic anisotropies help to decrease the error when adding the contribution of very small angular scales. In this right panel we also see that all the three scenarios give the same result  up to small angular scales (relative uncertainty of the order of 16\%): the reason is that the stochastic dipole is well above the corresponding instrumental noise level, see Fig.\,\ref{Nl}. Hence cosmic variance is the dominant source of uncertainty  in the  determination of the dipole.

		\medskip
		
\begin{figure}[htt!]
		\begin{center}
			%\captionsetup{justification=centering}
			\includegraphics[width=0.44\columnwidth]{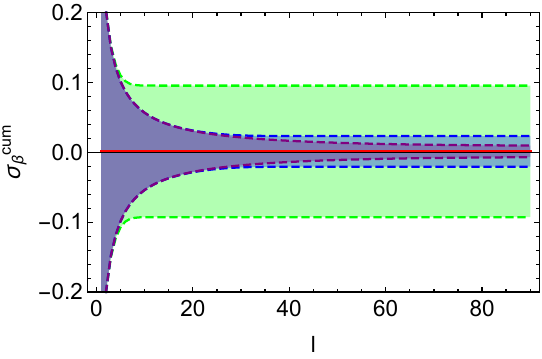}\qquad 	\includegraphics[width=0.44\columnwidth]{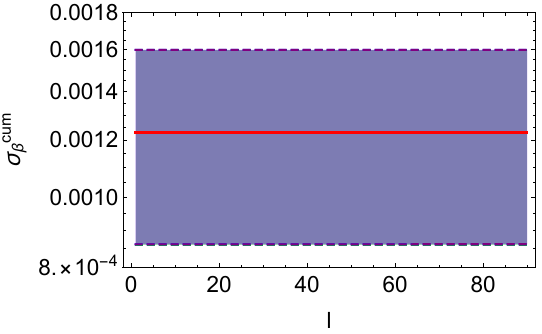}
			\caption{\small \label{BetaAstro} Variance on $\beta$ as function of angular resolution. The red line represents the value of the velocity reconstructed from CMB experiments. The shaded green, blue and purple areas correspond to the $1 \sigma$ region for the realistic scenario CE+ET and for the two futuristic scenarios described in the text (improvement in strain sensitivity of a factor 10 and 50 respectively). In the left panel we consider only the contribution of stochastic anisotropies, in the right one we also add the contribution from the non-stochastic dipole
			that plays an important role in controlling the error bars. The curves are indistinguishable up to very high multipoles, where the contribution of the stochastic component starts to be visible. We assume an integration time $T=10$ years.}
		\end{center}
	\end{figure}

{In Fig.\,\ref{BetaAstro} we represent the expected (cumulative)  precision  for constraining the velocity $\beta$ in each one of the scenarios under study, as a function of multipole $\ell$.  We show separately what can be achieved using stochastic anisotropies only (left panel), and adding the non-stochastic dipole (right panel).  Also in this case, we notice  that the non-stochastic dipole sets the size of the error bars. The three scenarios give the same result  up to small angular scales, with a relative uncertainty of the order of 30\%. }

		\medskip
		
\begin{figure}[htt!]
		\begin{center}
			%\captionsetup{justification=centering}
			\includegraphics[width=0.5\columnwidth]{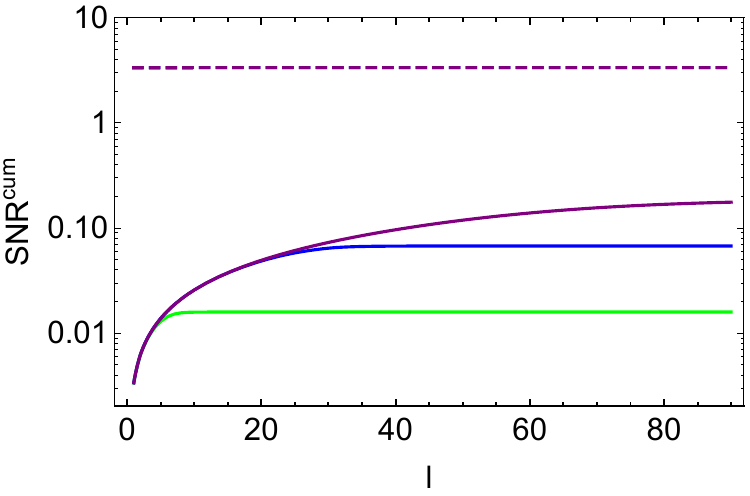}
			\caption{\small \label{SNRCum} Cumulative SNR for the three scenarios described in the text (the color code is the same as for figure \ref{SigmaCum}). With a solid line we represent the cumulative SNR associated to non-stochastic anisotropies, the dashed line also includes the contribution from the non-stochastic dipole, which dominates up to very small angular scales (the results for the three models are overlapped).  We assume an integration time $T=10$ years. }
		\end{center}
	\end{figure}

 In Fig.\,\ref{SNRCum} we show results for the cumulative SNR as a function of multipole, for each of the three scenarios under study. Also in this case, we plot separately the SNR associated to stochastic anisotropies alone, and then we add the contribution from the non-stochastic dipole (dashed lines). Each solid line reaches a plateau in correspondence to the angular scale at which the size of the  signal in figure \ref{Nl} reduces  below the instrumental noise level.  In this case too  we notice  that the non-stochastic dipole gives the dominant contribution to the cumulative signal to noise, up to very high multipoles, and that stochastic boost-induced anisotropies are not detectable alone in the realistic CE+ET scenario. 
 
 We stress that this analysis has only  the scope of illustrating our forecasting method. A realistic study of the detectability of the spectral density for an astrophysical background in the Hz band should include a more realistic noise and signal description, e.g. taking  into account a contribution from shot noise \cite{Jenkins:2019uzp, Jenkins:2019nks, Alonso:2020mva,Cusin:2019jpv}, and deviations from the simple power low behaviour in the spectral index due to the merging phase of the evolution of binaries, see e.g. \cite{Cusin:2019jpv}. A similar analysis can be done on the cross-correlation between the background energy density and the distribution of galaxies which, for an astrophysical background,  is expected to have an higher SNR that the auto-correlation -- see e.g. \cite{Cusin:2018rsq, Alonso:2020mva, Cusin:2019jpv,Yang:2020usq}.

%	
%	\textcolor{magenta}{For internal use: in the literature, to derive predictions for angular power spectrum of astrophysical background, we introduce the angular power spectrum of relative ansitropies as 
%\be
%C^{\text{rel}}_{\ell}=C_{\ell}\frac{(4\pi)^2}{\bar{\Omega}_{GW}^2}\,.
%\ee
%The reason of the factor $(4\pi)^2$ is that we want to have an integrated monopole \emph{per angular scale}. Notice that it is related to teh angulra power spectrum of $\Omega$ though factor of $(4\pi)^2$. We focus on astro background, where we know that typically $C_{\ell}^{\text{rel}}\sim 10^{-3}$ hence $C_{\ell}^{\Omega}\sim 6 \cdot 10^{-6}$. We fix a frequency of 25 Hz and we assume that $\Omega_{GW}(25 \text{Hz})\sim 10^{-9}$. Noise curves are computed referring to angular power spectrum of $\Omega_{GW}$. Then to have noise per multipole relative to the angular power spectrum of $\Omega_{GW}$, I need to divide the $N_{\ell}$ curve by factor $1/\Omega_{GW}^2$ (without factors of $(4\pi)^2$.) }

\section{Outlook}
\label{sec-outlook}
In this paper we studied kinematic effects associated with  the observer peculiar velocity on the energy density of the SGWB,
paying special attention to their dependence on the frequency slope  of the GW spectrum. We showed that a Doppler boost is responsible for the modulation and aberration of intrinsic anisotropies, and it additionally generates kinematic anisotropies from the rest-frame monopole of the SGWB. 
We provided analytic and ready-to-use expressions describing aberration, modulation and monopole-induced Doppler anisotropies
of the SGWB. We showed that  these effects are enhanced in the presence of large tilts of the frequency spectrum, and
examined explicit examples where these findings can be relevant.
We point out that a detection of boost-induced anisotropies can provide a complementary measurement of the spectral shape.  We outlined a method to forecast the precision with which the spectral shape can be measured by a given detector network.  For illustrative purposes, we applied this method to a simple case study: an extragalactic background  with spectral index $n_{\Omega}=2/3$ 
  induced by coalescing binaries
 detectable with ET+CE. In this simple case we find that the monopole-induced dipole will allow us to constrain the 
 spectral shape with a precision of about $16\%$.  {Interestingly, one can take a different perspective, and from the study of the kinematic dipole, try to extract information about our peculiar motion (with respect to the CMB rest frame), assuming that the spectral shape in a given frequency band is reconstructed via binning methods.  We show that a study of kinematic anisotropies with ET+CE can allow one to constrain our peculiar velocity with respect to the CMB frame with a precision of $30\%$.}

 Our analysis can be further developed in different directions. First, our forecasting method, which  we presented for the case of a power-law frequency dependence for  $\Omega_{\rm GW}$, can be  generalized and applied to any frequency  profile. It
 would be interesting to carry on such generalization to SGWB of cosmological origin  whose spectral index can have non-trivial dependence
 on frequencies. Moreover, while in section \ref{sec-prospects} we included only kinematic corrections at first order in the $\beta$-expansion, it
 would be 
  interesting to extend it at second  order in $\beta$, for better clarifying  consequences of 
  mixing
  among different multipoles.  At second order in $\beta$, one can  
  %\item Related with the previous point, while in section \ref{sec-prospects} we discussed prospects
  %for detecting $n_\Omega$ only, it would be interesting to
   %consider the possibility of 
   constrain
  additional parameters  controlling the properties of the spectrum, as for example $\alpha_\Omega$ which can be important
  for cosmological backgrounds (see e.g. Fig \ref{fig:plot1}).

  % Finally, it would be interesting   to use
 %Doppler-induced anisotropies  for  distinguishing whether a background from black hole mergers has primordial (e.g. PBH) or astrophysical origin. Indeed typically in the Hz band $n_{\Omega}=2/3$ for an astrophysical background while typically $n_{\Omega}<0$ for a population of PBH with merger rate $\mathcal{R}\propto (1+z)^{\alpha}$ and $\alpha \in [0, 10]$. (See e.g. \cite{Sasaki_2018, Carr_2016} for recent reviews on PBH.) %\textcolor{magenta}{Give references XXX}.
  A reconstruction of the spectral shape is a precious tool to distinguish background components with different origins. Current reconstruction  methods require a detection in multiple frequency bands (binning), via small-band searches, which are more challenging than the broad band searches currently implemented by the LIGO-Virgo collaboration, % \textcolor{magenta}{give references}, 
   in particular if the signal has a steep spectral shape, with a small frequency bin dominating the total SNR budget. The method that we present in this work may allow one to extract information on the spectral shape without the need of multiple band reconstruction.   
   
   {Finally, SGWB kinematic dipole is a new observable which depends on both the spectral shape and on the velocity with which we move with respect to the emission frame (traditionally identified with the CMB rest frame). This new observable provides us with an independent way of reconstructing our kinematc motion: studying it could potentially shed light on the existing discrepancy in the value of $\beta$ reconstructed from CMB and from galaxy number counts (see e.g. \cite{Dalang:2021ruy} and references therein for a recent critical analysis). 
   %Applications to realistic scenarios of mixed population of PBH and AGWB will be presented in a future work.  
 We hope to return to these topics soon. }

% \noindent
 %{\bf \color{red} Things to further discuss/ develop} %in the future:}
 
 \smallskip

\subsection*{Acknowledgments}
It is a pleasure to thank Nicola Bartolo, Robert Caldwell, Raphael Flauger, Marco Peloso,  Angelo Ricciardone and Lorenzo Sorbo for useful discussions. GC  is funded by Swiss National Science Foundation (Ambizione Grant). 
GT is partially funded by the STFC grant ST/T000813/1.   

\begin{appendix}
%\appendix

\section{Products of spherical harmonics and the Gaunt coefficient}\label{lm}

In this appendix we summarise some properties of integrals over products of three spherical harmonics, and associated quantities. It is based on appendix H of \cite{Cusin:2016kqx}. The key result is that the integral of three spin-weighted spherical harmonics can be written as
\begin{align}\label{1111}
\int d\Omega \,_{s_1}&Y_{\ell_1 m_1}\,_{s_2}Y_{\ell_2 m_2}\,_{s_3}Y_{\ell_3 m_3}=\nn\\
&=\sqrt{\frac{(2\ell_1+1)(2\ell_2+1)(2\ell_3+1)}{4\pi}}\left(
\begin{array}{ccc}
\ell_1&\ell_2&\ell_3\\
-s_1&-s_2&-s_3
\end{array}
\right)\left(
\begin{array}{ccc}
\ell_1&\ell_2&\ell_3\\
m_1&m_2&m_3
\end{array}
\right)\,.
\end{align}
The $3-j$ symbols that appear in this expression satisfy the following properties
\begin{align}\label{11}
\left(
\begin{array}{ccc}
\ell_1&\ell_2&\ell_3\\
m_1&m_2&m_3\\
\end{array}
\right)&=
\left(
\begin{array}{ccc}
\ell_2&\ell_3&\ell_1\\
m_2&m_3&m_1\\
\end{array}
\right)=
\left(
\begin{array}{ccc}
\ell_3&\ell_1&\ell_2\\
m_3&m_1&m_2\\
\end{array}
\right)\\
&=(-)^{\ell_1+\ell_2+\ell_3}
\left(
\begin{array}{ccc}
\ell_1&\ell_3&\ell_2\\
m_1&m_3&m_2\\
\end{array}
\right)\\
&=(-)^{\ell_1+\ell_2+\ell_3}
\left(
\begin{array}{ccc}
\ell_1&\ell_2&\ell_3\\
-m_1&-m_2&-m_3\\
\end{array}\label{111}
\right)\,.
\end{align}
Specifically, they are identically zero whenever any of the following conditions are violated
\be
m_1+m_2+m_3=0\,,\qquad |\ell_i-\ell_j|\leq \ell_k\leq \ell_i+\ell_j\,,\qquad \{i\,,j\}=\{1,2,3\}\,.
\ee

Some important quantities that are used in this paper are defined as
\begin{eqnarray}
\mathcal{W}^{m_1 m_2 m_3}_{\ell_1 \ell_2 \ell_3} &\equiv& \int \dd \Omega\,
Y^{\star}_{\ell_1 m_1} Y_{\ell_2 m_2} Y_{\ell_3 m_3}\,,
\end{eqnarray}
This is effectively the Gaunt coefficient (up to the complex conjugation of the first spherical harmonic which leads to some sign changes). It is given by
\begin{align}
\mathcal{W}_{\ell_1 \ell_2 \ell_3}^{m_1 m_2 m_3} &=
(-1)^{m_1}\troisj{\ell_1}{\ell_2}{\ell_3}{-m_1}{m_2}{m_3}\,{\cal F}_{\ell_1\ell_2\ell_3}\,,\label{CC}\\
{\cal F}_{\ell \ell_1 \ell_2} &= \sqrt{\frac{(2\ell+1)(2
    \ell_1+1)(2 \ell_2+2)}{4\pi}}\troisj{\ell}{\ell_1}{\ell_2}{0}{0}{0}\, .
\end{align}
Another useful relation is that 
\be
\mathcal{I}_{\ell_1 \ell_2 \ell_3}^{m_1 m_2 m_3} =\frac{1}{2}\left[\ell_3(\ell_3+1)-\ell_2(\ell_2+1)-\ell_1(\ell_1+1)\right]\mathcal{W}_{\ell_1 \ell_2 \ell_3}^{m_1 m_2 m_3}\,,
\ee
where
\be
\mathcal{I}_{\ell_1 \ell_2 \ell_3}^{m_1 m_2 m_3}\equiv \int \dd \Omega\,
Y^{\star}_{\ell_1 m_1} \nabla^a Y_{\ell_2 m_2} \nabla_a Y_{\ell_3 m_3}\,. 
\ee

%{\color{red}
%\section{References to add along the way}}

%\begin{itemize}
%\item
%Bounds on anisotropies and SGWB: 
%\cite{Abbott:2021xxi,Abbott:2021jel}
%\item slope in the IR is discussed in . Explicit paper where drops are further analyzed \cite{Inomata:2019ivs}.  
%Moreover, \cite{Espinosa:2018eve,Kohri:2018awv,Cai:2019amo}.
%\item Renaux Petel \cite{Fumagalli:2020nvq} discusses phenomena of particle production and rapid turns, which
%can lead to oscillatory OmegaGW that should amplify its spectral index.
%\end{itemize}

{
\section{On the validity of the $\beta$ expansion in section \ref{sec-ex-infl}}
\label{app_issuexp}
In this  Appendix we reconsider the model of section \ref{sec-ex-infl}, 
to demonstrate the
consistency of 
 %show
 %that
  a  $\beta^2$
   truncation in the expansion %of eq \eqref{} 
   as 
 in  
formulas \eqref{monanis}-\eqref{quapanis}.
% is  consistent for the case under 
%consideration.
  In fact, given that the
spectral tilts are large  -- see Fig \ref{fig:plot1} --  we
should be cautious, and  check whether contributions weighted
by  higher powers of $\beta$   invalidate or not our formulas.
 %
%are larger than the ones we consider, invalidating our formulas truncated at second order in a $\beta$ expansion. 
 %
We start from formula \eqref{genexpom4} and expand up   fourth order in a $\beta$
expansion. We obtain the following generalization of eqs \eqref{monanis}-\eqref{quapanis}: 
\bea
\label{monanisA}
M(f)&=&\frac{\beta^2}{6} \left( 8+n_\Omega \left( n_\Omega-6\right)
+\alpha_\Omega
\right)
\nonumber
\\
&&+
\frac{\beta^4}{24}
\left(2 n_\Omega^3 
-23 n_\Omega^2-23 \alpha_\Omega
+ 94 n_\Omega  
+6 \alpha_\Omega\,n_\Omega
+2 \gamma_\Omega
-136
\right)
\,,
\\
\label{dipanisA}
D(f)&=& \beta \left(4-n_\Omega\right)
+\beta^3 \left(4 n_\Omega-8-\frac{n_\Omega^2}{2}
-\frac{\alpha_\Omega}{2}
 \right)
\,,
\\
\label{quapanisA}
Q(f)&=&\beta^2\left(10-\frac{9 n_\Omega}{2} +\frac{n_\Omega^2}{2}+\frac{\alpha_\Omega}{2}\right)
\nonumber
\\
&&+ \frac{\beta^4}{4} 
\left( n_\Omega^3 
-13 n_\Omega^2-13 \alpha_\Omega
+ 56 n_\Omega  
+3 \alpha_\Omega\,n_\Omega
+ \gamma_\Omega
-80
\right)
\,,
\eea
indicating respectively the monopole, dipole, quadrupole  Doppler contributions
expanded up to $\beta^4$. 
We define $\gamma_\Omega\,\equiv\,d \alpha_\Omega/d \ln f$.
The
monopole and quadrupole are sensitive to the fourth power of $\beta$,
the dipole to the third power only.

\begin{figure}[h!]
\centering
    \includegraphics[width = 0.32 \textwidth]{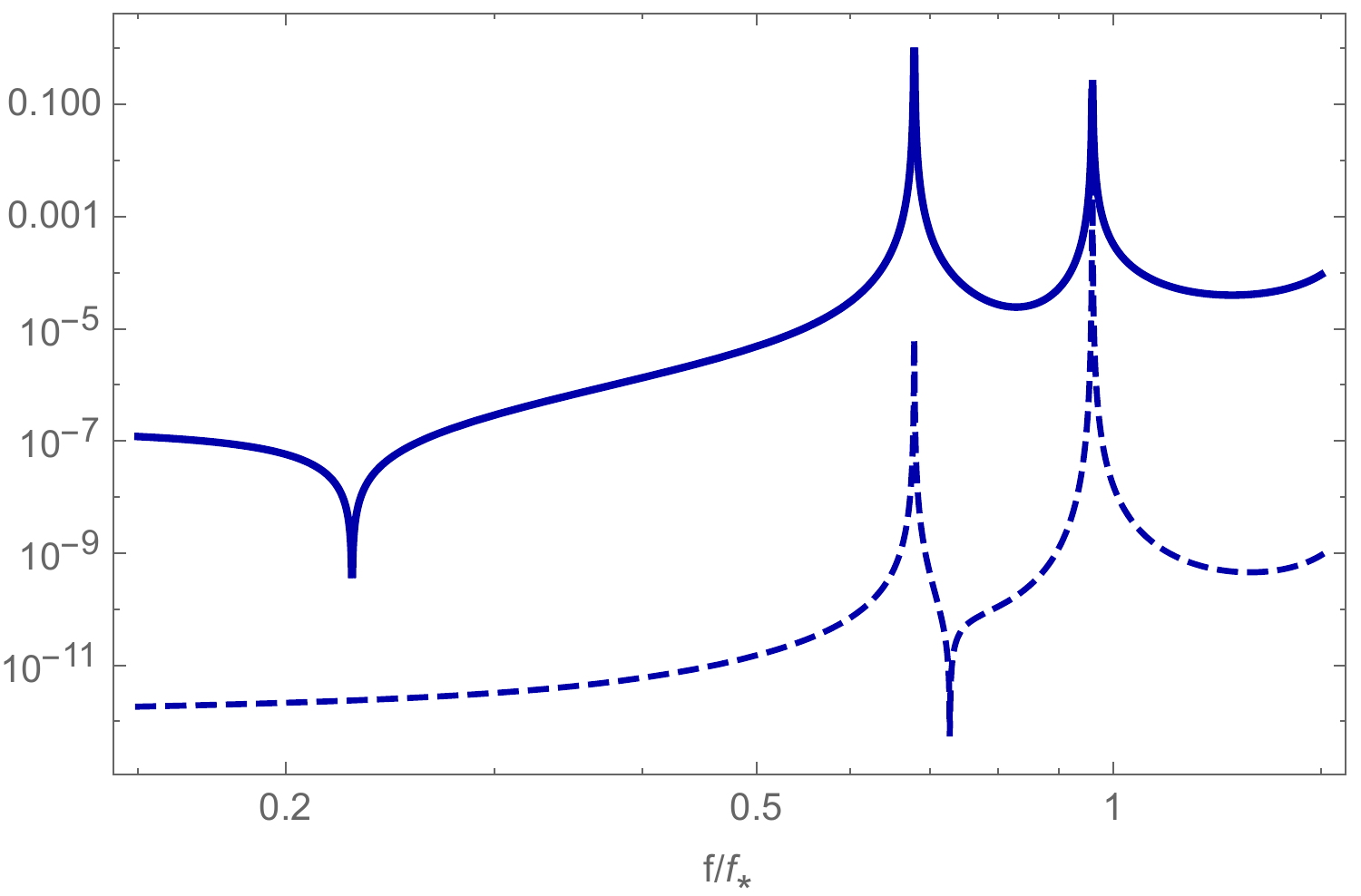}
        \includegraphics[width = 0.32 \textwidth]{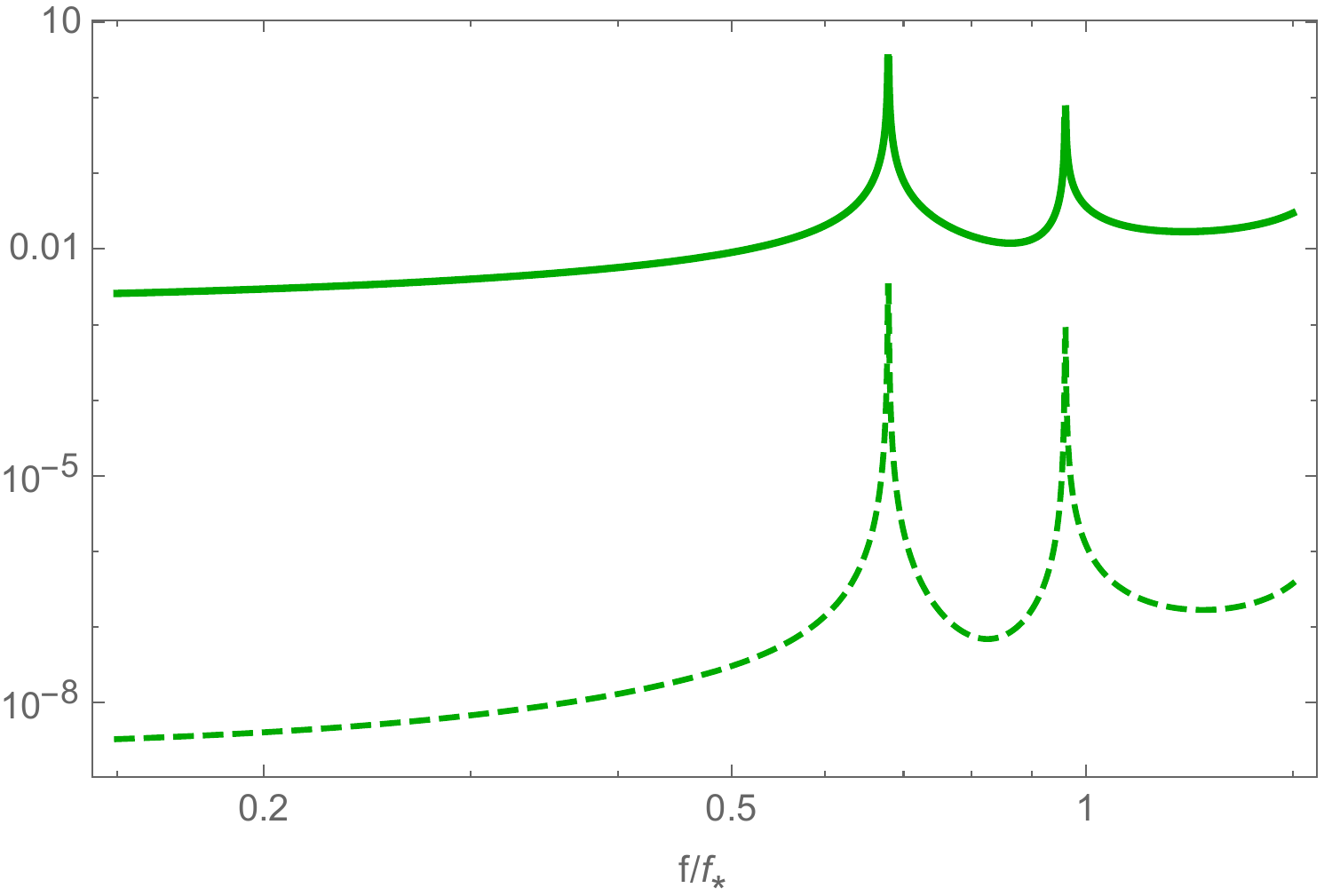}
                \includegraphics[width = 0.32 \textwidth]{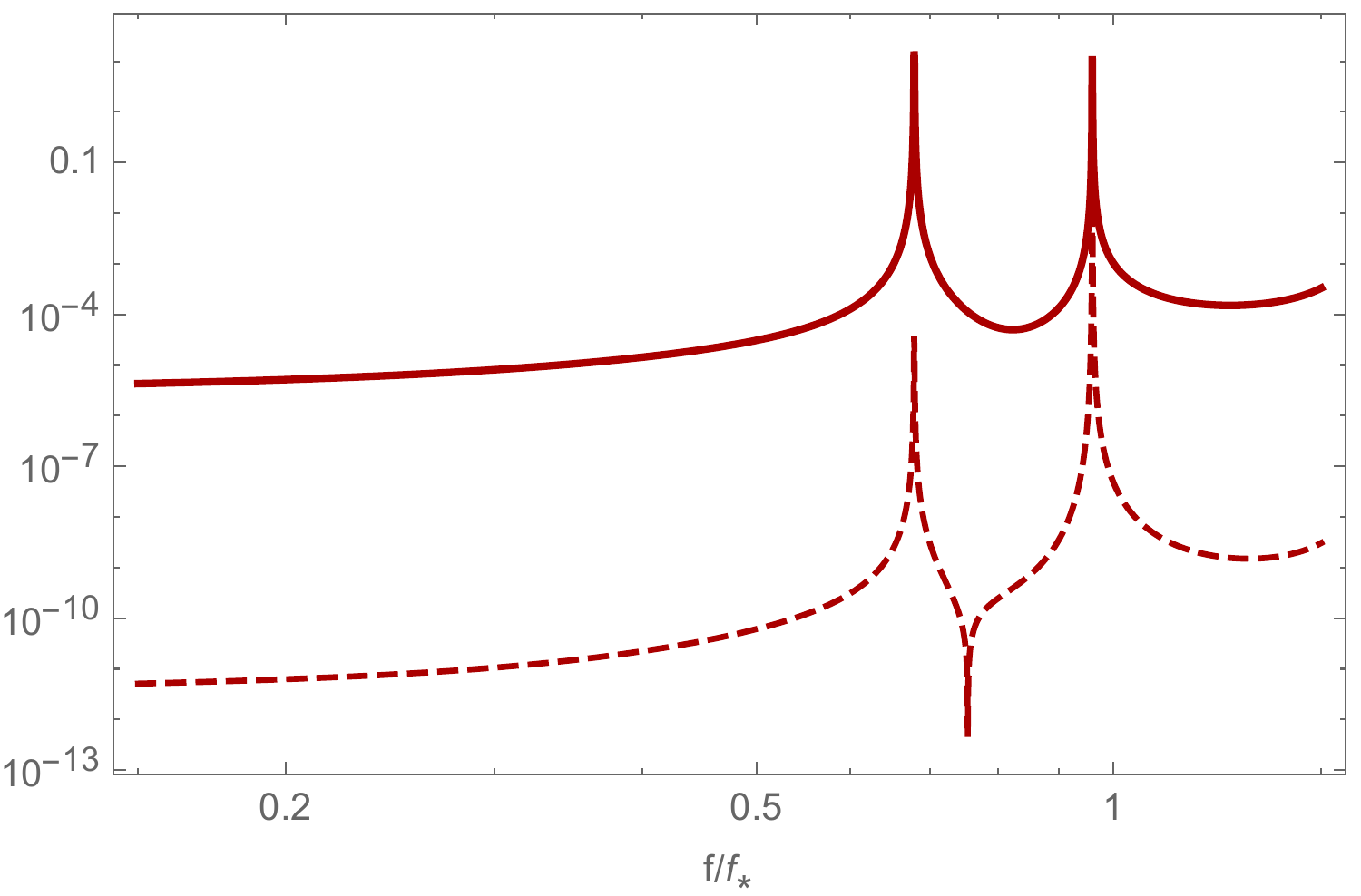}
 \caption{\small {
Kinematic contributions to the  monopole ($M(f)$, left) dipole ($D(f)$, middle), and quadrupole ($Q(f)$, right) anisotropies,  computed using formulas expanded 
up to order $\beta^2$ (continuous lines),  together with the difference between  their values computed up to  order $\beta^2$ and up to  order $\beta^4$ (dashed lines).   We use the same values
of parameters of Fig \ref{fig:plot5}.}
  }
 \label{fig:plot5aa}
\end{figure}

In Fig \ref{fig:plot5aa} we represent the frequency dependence of the kinematic anisotropies,  computed using formulas expanded 
up to order $\beta^2$ in eqs \eqref{monanis}-\eqref{quapanis},  together with the difference between  their values computed up to order $\beta^2$ and up to  order $\beta^4$
using eqs \eqref{monanisA}-\eqref{quapanisA}.  The difference is almost always orders of magnitude smaller
than the  kinematic anisotropies computed at order $\beta^2$, except at
 the  frequency $f/f_\star\,=\,2/\sqrt{3}$. However, 
 as explained  
in footnote 
 \ref{foot_discl2}, 
 for that precise frequency   resonant effects  are expected to  be smoothed out,
 at least  in realistic situations. Hence, we conclude that the expansions used for making the plot in Fig \ref{fig:plot5} are sufficient for
 the values of the parameters chosen, and the results of section \ref{sec-ex-infl} are
 reliable.
}

\end{appendix}

\providecommand{\href}[2]{#2}\begingroup\raggedright\endgroup


\begin{thebibliography}{10}

\bibitem{Regimbau:2011rp}
T.~Regimbau, ``{The astrophysical gravitational wave stochastic background},''
  \href{http://dx.doi.org/10.1088/1674-4527/11/4/001}{{\em Res. Astron.
  Astrophys.} {\bfseries 11} (2011) 369--390},
\href{http://arxiv.org/abs/1101.2762}{{\ttfamily arXiv:1101.2762
  [astro-ph.CO]}}.
%%CITATION = ARXIV:1101.2762;%%.

\bibitem{Caprini:2018mtu}
C.~Caprini and D.~G. Figueroa, ``{Cosmological Backgrounds of Gravitational
  Waves},'' \href{http://dx.doi.org/10.1088/1361-6382/aac608}{{\em Class.
  Quant. Grav.} {\bfseries 35} no.~16, (2018) 163001},
\href{http://arxiv.org/abs/1801.04268}{{\ttfamily arXiv:1801.04268
  [astro-ph.CO]}}.
%%CITATION = ARXIV:1801.04268;%%.

\bibitem{Maggiore:2018sht}
M.~Maggiore, {\em {Gravitational Waves. Vol. 2: Astrophysics and Cosmology}}.
\newblock Oxford University Press,
2018.
\newblock
%%CITATION = INSPIRE-1664982;%%.

\bibitem{Abbott:2021xxi}
{\bfseries LIGO Scientific, Virgo, KAGRA} Collaboration, R.~Abbott {\em
  et~al.}, ``{Upper Limits on the Isotropic Gravitational-Wave Background from
  Advanced LIGO's and Advanced Virgo's Third Observing Run},''
  \href{http://arxiv.org/abs/2101.12130}{{\ttfamily arXiv:2101.12130 [gr-qc]}}.

\bibitem{TheLIGOScientific:2016xzw}
{\bfseries Virgo, LIGO Scientific} Collaboration, B.~P. Abbott {\em et~al.},
  ``{Directional Limits on Persistent Gravitational Waves from Advanced LIGO?s
  First Observing Run},''
  \href{http://dx.doi.org/10.1103/PhysRevLett.118.121102}{{\em Phys. Rev.
  Lett.} {\bfseries 118} no.~12, (2017) 121102},
\href{http://arxiv.org/abs/1612.02030}{{\ttfamily arXiv:1612.02030 [gr-qc]}}.
%%CITATION = ARXIV:1612.02030;%%.

\bibitem{Abbott:2021jel}
{\bfseries LIGO Scientific, Virgo, KAGRA} Collaboration, R.~Abbott {\em
  et~al.}, ``{Search for anisotropic gravitational-wave backgrounds using data
  from Advanced LIGO's and Advanced Virgo's first three observing runs},''
  \href{http://arxiv.org/abs/2103.08520}{{\ttfamily arXiv:2103.08520 [gr-qc]}}.

\bibitem{Arzoumanian:2020vkk}
{\bfseries NANOGrav} Collaboration, Z.~Arzoumanian {\em et~al.}, ``{The
  NANOGrav 12.5 yr Data Set: Search for an Isotropic Stochastic
  Gravitational-wave Background},''
  \href{http://dx.doi.org/10.3847/2041-8213/abd401}{{\em Astrophys. J. Lett.}
  {\bfseries 905} no.~2, (2020) L34},
\href{http://arxiv.org/abs/2009.04496}{{\ttfamily arXiv:2009.04496
  [astro-ph.HE]}}.
%%CITATION = ARXIV:2009.04496;%%.

\bibitem{Smoot:1977bs}
G.~F. Smoot, M.~V. Gorenstein, and R.~A. Muller, ``{Detection of Anisotropy in
  the Cosmic Black Body Radiation},''
\href{http://dx.doi.org/10.1103/PhysRevLett.39.898}{{\em Phys. Rev. Lett.}
  {\bfseries 39} (1977) 898}.
%%CITATION = PRLTA,39,898;%%.

\bibitem{Kogut:1993ag}
A.~Kogut {\em et~al.}, ``{Dipole anisotropy in the COBE DMR first year sky
  maps},'' \href{http://dx.doi.org/10.1086/173453}{{\em Astrophys. J.}
  {\bfseries 419} (1993) 1},
\href{http://arxiv.org/abs/astro-ph/9312056}{{\ttfamily arXiv:astro-ph/9312056
  [astro-ph]}}.
%%CITATION = ASTRO-PH/9312056;%%.

\bibitem{Bennett:2003bz}
{\bfseries WMAP} Collaboration, C.~L. Bennett {\em et~al.}, ``{First year
  Wilkinson Microwave Anisotropy Probe (WMAP) observations: Preliminary maps
  and basic results},'' \href{http://dx.doi.org/10.1086/377253}{{\em Astrophys.
  J. Suppl.} {\bfseries 148} (2003) 1--27},
\href{http://arxiv.org/abs/astro-ph/0302207}{{\ttfamily arXiv:astro-ph/0302207
  [astro-ph]}}.
%%CITATION = ASTRO-PH/0302207;%%.

\bibitem{Aghanim:2013suk}
{\bfseries Planck} Collaboration, N.~Aghanim {\em et~al.}, ``{Planck 2013
  results. XXVII. Doppler boosting of the CMB: Eppur si muove},''
  \href{http://dx.doi.org/10.1051/0004-6361/201321556}{{\em Astron. Astrophys.}
  {\bfseries 571} (2014) A27},
\href{http://arxiv.org/abs/1303.5087}{{\ttfamily arXiv:1303.5087
  [astro-ph.CO]}}.
%%CITATION = ARXIV:1303.5087;%%.

\bibitem{Henry:1969im}
G.~R. Henry, R.~B. Feduniak, J.~E. Silver, and M.~A. Peterson, ``{Distribution
  of black body cavity radiation in a moving frame of reference},''
\href{http://dx.doi.org/10.1103/PhysRev.176.1451}{{\em Phys. Rev.} {\bfseries
  176} (1968) 1451--1455}.
%%CITATION = PHRVA,176,1451;%%.

\bibitem{Peebles:1968zz}
P.~J.~E. Peebles and D.~T. Wilkinson, ``{Comment on the Anisotropy of the
  Primeval Fireball},''
\href{http://dx.doi.org/10.1103/PhysRev.174.2168}{{\em Phys. Rev.} {\bfseries
  174} (1968) 2168--2168}.
%%CITATION = PHRVA,174,2168;%%.

\bibitem{Gorski:1990ua}
K.~M. Gorski, ``{Will COBE challenge the inflationary paradigm? 1. Cosmic
  microwave background anisotropies versus large scale streaming motions
  (Revised)},''
{\em Submitted to: Ap. J.(Letters) 1990 July XX} (1990) .
%%CITATION = LA-UR-90-40;%%.

\bibitem{CvL02}
A.~Challinor and F.~van Leeuwen, ``{Peculiar velocity effects in high
  resolution microwave background experiments},''
  \href{http://dx.doi.org/10.1103/PhysRevD.65.103001}{{\em Phys. Rev.}
  {\bfseries D65} (2002) 103001},
\href{http://arxiv.org/abs/astro-ph/0112457}{{\ttfamily arXiv:astro-ph/0112457
  [astro-ph]}}.
%%CITATION = ASTRO-PH/0112457;%%.

\bibitem{Menzies:2004vr}
D.~Menzies and G.~J. Mathews, ``{Peculiar velocity and deaberration of the
  sky},'' \href{http://dx.doi.org/10.1086/428936}{{\em Astrophys. J.}
  {\bfseries 624} (2005) 7--9},
\href{http://arxiv.org/abs/astro-ph/0409175}{{\ttfamily arXiv:astro-ph/0409175
  [astro-ph]}}.
%%CITATION = ASTRO-PH/0409175;%%.

\bibitem{BR06}
S.~Burles and S.~Rappaport, ``{Aberration of the Cosmic Microwave
  Background},'' \href{http://dx.doi.org/10.1086/503743}{{\em Astrophys. J.}
  {\bfseries 641} (2006) L1},
\href{http://arxiv.org/abs/astro-ph/0601559}{{\ttfamily arXiv:astro-ph/0601559
  [astro-ph]}}.
%%CITATION = ASTRO-PH/0601559;%%.

\bibitem{Kosowsky:2010jm}
A.~Kosowsky and T.~Kahniashvili, ``{The Signature of Proper Motion in the
  Microwave Sky},''
  \href{http://dx.doi.org/10.1103/PhysRevLett.106.191301}{{\em Phys. Rev.
  Lett.} {\bfseries 106} (2011) 191301},
\href{http://arxiv.org/abs/1007.4539}{{\ttfamily arXiv:1007.4539
  [astro-ph.CO]}}.
%%CITATION = ARXIV:1007.4539;%%.

\bibitem{Amendola:2010ty}
L.~Amendola {\em et~al.}, ``{Measuring our peculiar velocity on the CMB with
  high-multipole off-diagonal correlations},''
  \href{http://dx.doi.org/10.1088/1475-7516/2011/07/027}{{\em JCAP} {\bfseries
  1107} (2011) 027},
\href{http://arxiv.org/abs/1008.1183}{{\ttfamily arXiv:1008.1183
  [astro-ph.CO]}}.
%%CITATION = ARXIV:1008.1183;%%.

%\cite{Mukherjee:2013zbi}
\bibitem{Mukherjee:2013zbi}
S.~Mukherjee, A.~De and T.~Souradeep,
``{Statistical isotropy violation of CMB Polarization sky due to Lorentz boost},''
%Phys. Rev. D \textbf{89} (2014) no.8, 083005
%doi:10.1103/PhysRevD.89.083005
\href{http://arxiv.org/abs/1309.3800}{{\ttfamily arXiv:1309.3800
  [astro-ph.CO]}}.
%[arXiv:1309.3800 [astro-ph.CO]].
%17 citations counted in INSPIRE as of 02 Feb 2022

\bibitem{Bonvin:2006en}
C.~Bonvin, R.~Durrer, and M.~Kunz, ``{The dipole of the luminosity distance: a
  direct measure of H(z)},''
  \href{http://dx.doi.org/10.1103/PhysRevLett.96.191302}{{\em Phys. Rev. Lett.}
  {\bfseries 96} (2006) 191302},
\href{http://arxiv.org/abs/astro-ph/0603240}{{\ttfamily arXiv:astro-ph/0603240
  [astro-ph]}}.
%%CITATION = ASTRO-PH/0603240;%%.

\bibitem{10.1093/mnras/206.2.377}
G.~F.~R. Ellis and J.~E. Baldwin, ``{On the expected anisotropy of radio source
  counts},'' {\em Monthly Notices of the Royal Astronomical Society} {\bfseries
  206} no.~2, (01, 1984) 377--381.

\bibitem{Maartens:2017qoa}
R.~Maartens, C.~Clarkson, and S.~Chen, ``{The kinematic dipole in galaxy
  redshift surveys},''
  \href{http://dx.doi.org/10.1088/1475-7516/2018/01/013}{{\em JCAP} {\bfseries
  1801} (2018) 013},
\href{http://arxiv.org/abs/1709.04165}{{\ttfamily arXiv:1709.04165
  [astro-ph.CO]}}.
%%CITATION = ARXIV:1709.04165;%%.

\bibitem{Pant:2018smd}
N.~Pant, A.~Rotti, C.~A.~P. Bengaly, and R.~Maartens, ``{Measuring our velocity
  from fluctuations in number counts},''
  \href{http://dx.doi.org/10.1088/1475-7516/2019/03/023}{{\em JCAP} {\bfseries
  1903} (2019) 023},
\href{http://arxiv.org/abs/1808.09743}{{\ttfamily arXiv:1808.09743
  [astro-ph.CO]}}.
%%CITATION = ARXIV:1808.09743;%%.

\bibitem{Bartolo:2022pez}
N.~Bartolo {\em et~al.}, ``{Probing Anisotropies of the Stochastic
  Gravitational Wave Background with LISA},''
  \href{http://arxiv.org/abs/2201.08782}{{\ttfamily arXiv:2201.08782
  [astro-ph.CO]}}.

\bibitem{Jenkins:2018lvb}
A.~C. Jenkins and M.~Sakellariadou, ``{Anisotropies in the stochastic
  gravitational-wave background: Formalism and the cosmic string case},''
  \href{http://dx.doi.org/10.1103/PhysRevD.98.063509}{{\em Phys. Rev.}
  {\bfseries D98} no.~6, (2018) 063509},
\href{http://arxiv.org/abs/1802.06046}{{\ttfamily arXiv:1802.06046
  [astro-ph.CO]}}.
%%CITATION = ARXIV:1802.06046;%%.

\bibitem{mckinley}
J.~McKinley, ``{Relativistic transformations of light power},''
\href{http://dx.doi.org/...}{{\em Am. J. Phys.} {\bfseries 47} (1979) 602}.
%%CITATION = PHRVA,174,2168;%%.

\bibitem{Landau:1987gn}
L.~D. Landau and E.~M. Lifshitz, {\em {Course of Theoretical Physics, Vol. II:
  Classical field theory}}.
\newblock
1987.
\newblock
%%CITATION = INSPIRE-256560;%%.

\bibitem{Kamionkowski:2002nd}
M.~Kamionkowski and L.~Knox, ``{Aspects of the cosmic microwave background
  dipole},'' \href{http://dx.doi.org/10.1103/PhysRevD.67.063001}{{\em Phys.
  Rev.} {\bfseries D67} (2003) 063001},
\href{http://arxiv.org/abs/astro-ph/0210165}{{\ttfamily arXiv:astro-ph/0210165
  [astro-ph]}}.
%%CITATION = ASTRO-PH/0210165;%%.

\bibitem{Allen:1996gp}
B.~Allen and A.~C. Ottewill, ``{Detection of anisotropies in the gravitational
  wave stochastic background},''
  \href{http://dx.doi.org/10.1103/PhysRevD.56.545}{{\em Phys. Rev.} {\bfseries
  D56} (1997) 545--563},
\href{http://arxiv.org/abs/gr-qc/9607068}{{\ttfamily arXiv:gr-qc/9607068
  [gr-qc]}}.
%%CITATION = GR-QC/9607068;%%.

\bibitem{Bartolo:2016ami}
N.~Bartolo {\em et~al.}, ``{Science with the space-based interferometer LISA.
  IV: Probing inflation with gravitational waves},''
  \href{http://dx.doi.org/10.1088/1475-7516/2016/12/026}{{\em JCAP} {\bfseries
  1612} (2016) 026},
\href{http://arxiv.org/abs/1610.06481}{{\ttfamily arXiv:1610.06481
  [astro-ph.CO]}}.
%%CITATION = ARXIV:1610.06481;%%.

\bibitem{Cannone:2014uqa}
D.~Cannone, G.~Tasinato, and D.~Wands, ``{Generalised tensor fluctuations and
  inflation},'' \href{http://dx.doi.org/10.1088/1475-7516/2015/01/029}{{\em
  JCAP} {\bfseries 01} (2015) 029},
  \href{http://arxiv.org/abs/1409.6568}{{\ttfamily arXiv:1409.6568
  [astro-ph.CO]}}.

\bibitem{Bartolo:2015qvr}
N.~Bartolo, D.~Cannone, A.~Ricciardone, and G.~Tasinato, ``{Distinctive
  signatures of space-time diffeomorphism breaking in EFT of inflation},''
  \href{http://dx.doi.org/10.1088/1475-7516/2016/03/044}{{\em JCAP} {\bfseries
  03} (2016) 044}, \href{http://arxiv.org/abs/1511.07414}{{\ttfamily
  arXiv:1511.07414 [astro-ph.CO]}}.

\bibitem{Ricciardone:2016lym}
A.~Ricciardone and G.~Tasinato, ``{Primordial gravitational waves in supersolid
  inflation},'' \href{http://dx.doi.org/10.1103/PhysRevD.96.023508}{{\em Phys.
  Rev. D} {\bfseries 96} no.~2, (2017) 023508},
  \href{http://arxiv.org/abs/1611.04516}{{\ttfamily arXiv:1611.04516
  [astro-ph.CO]}}.

\bibitem{Matarrese:1992rp}
S.~Matarrese, O.~Pantano, and D.~Saez, ``{A General relativistic approach to
  the nonlinear evolution of collisionless matter},''
\href{http://dx.doi.org/10.1103/PhysRevD.47.1311}{{\em Phys. Rev.} {\bfseries
  D47} (1993) 1311--1323}.
%%CITATION = PHRVA,D47,1311;%%.

\bibitem{Matarrese:1993zf}
S.~Matarrese, O.~Pantano, and D.~Saez, ``{General relativistic dynamics of
  irrotational dust: Cosmological implications},''
  \href{http://dx.doi.org/10.1103/PhysRevLett.72.320}{{\em Phys. Rev. Lett.}
  {\bfseries 72} (1994) 320--323},
\href{http://arxiv.org/abs/astro-ph/9310036}{{\ttfamily arXiv:astro-ph/9310036
  [astro-ph]}}.
%%CITATION = ASTRO-PH/9310036;%%.

\bibitem{Matarrese:1997ay}
S.~Matarrese, S.~Mollerach, and M.~Bruni, ``{Second order perturbations of the
  Einstein-de Sitter universe},''
  \href{http://dx.doi.org/10.1103/PhysRevD.58.043504}{{\em Phys. Rev.}
  {\bfseries D58} (1998) 043504},
\href{http://arxiv.org/abs/astro-ph/9707278}{{\ttfamily arXiv:astro-ph/9707278
  [astro-ph]}}.
%%CITATION = ASTRO-PH/9707278;%%.

\bibitem{Noh:2004bc}
H.~Noh and J.-c. Hwang, ``{Second-order perturbations of the Friedmann world
  model},''
\href{http://dx.doi.org/10.1103/PhysRevD.69.104011}{{\em Phys. Rev.} {\bfseries
  D69} (2004) 104011}.
%%CITATION = PHRVA,D69,104011;%%.

\bibitem{Nakamura:2004rm}
K.~Nakamura, ``{Second-order gauge invariant cosmological perturbation theory:
  Einstein equations in terms of gauge invariant variables},''
  \href{http://dx.doi.org/10.1143/PTP.117.17}{{\em Prog. Theor. Phys.}
  {\bfseries 117} (2007) 17--74},
\href{http://arxiv.org/abs/gr-qc/0605108}{{\ttfamily arXiv:gr-qc/0605108
  [gr-qc]}}.
%%CITATION = GR-QC/0605108;%%.

\bibitem{Ananda:2006af}
K.~N. Ananda, C.~Clarkson, and D.~Wands, ``{The Cosmological gravitational wave
  background from primordial density perturbations},''
  \href{http://dx.doi.org/10.1103/PhysRevD.75.123518}{{\em Phys. Rev.}
  {\bfseries D75} (2007) 123518},
\href{http://arxiv.org/abs/gr-qc/0612013}{{\ttfamily arXiv:gr-qc/0612013
  [gr-qc]}}.
%%CITATION = GR-QC/0612013;%%.

\bibitem{Baumann:2007zm}
D.~Baumann, P.~J. Steinhardt, K.~Takahashi, and K.~Ichiki, ``{Gravitational
  Wave Spectrum Induced by Primordial Scalar Perturbations},''
  \href{http://dx.doi.org/10.1103/PhysRevD.76.084019}{{\em Phys. Rev.}
  {\bfseries D76} (2007) 084019},
\href{http://arxiv.org/abs/hep-th/0703290}{{\ttfamily arXiv:hep-th/0703290
  [hep-th]}}.
%%CITATION = HEP-TH/0703290;%%.

\bibitem{Sasaki_2018}
M.~Sasaki, T.~Suyama, T.~Tanaka, and S.~Yokoyama, ``Primordial black
  holes?perspectives in gravitational wave astronomy,''
  \href{http://dx.doi.org/10.1088/1361-6382/aaa7b4}{{\em Classical and Quantum
  Gravity} {\bfseries 35} no.~6, (Feb, 2018) 063001}.
  \url{http://dx.doi.org/10.1088/1361-6382/aaa7b4}.

\bibitem{Carr_2016}
B.~Carr, F.~Kühnel, and M.~Sandstad, ``Primordial black holes as dark
  matter,'' \href{http://dx.doi.org/10.1103/physrevd.94.083504}{{\em Physical
  Review D} {\bfseries 94} no.~8, (Oct, 2016) }.
  \url{http://dx.doi.org/10.1103/PhysRevD.94.083504}.

\bibitem{Saito:2008jc}
R.~Saito and J.~Yokoyama, ``{Gravitational wave background as a probe of the
  primordial black hole abundance},''
  \href{http://dx.doi.org/10.1103/PhysRevLett.102.161101,
  10.1103/PhysRevLett.107.069901}{{\em Phys. Rev. Lett.} {\bfseries 102} (2009)
  161101}, \href{http://arxiv.org/abs/0812.4339}{{\ttfamily arXiv:0812.4339
  [astro-ph]}}.
[Erratum: Phys. Rev. Lett.107,069901(2011)].
%%CITATION = ARXIV:0812.4339;%%.

\bibitem{Saito:2009jt}
R.~Saito and J.~Yokoyama, ``{Gravitational-Wave Constraints on the Abundance of
  Primordial Black Holes},'' \href{http://dx.doi.org/10.1143/PTP.126.351,
  10.1143/PTP.123.867}{{\em Prog. Theor. Phys.} {\bfseries 123} (2010)
  867--886}, \href{http://arxiv.org/abs/0912.5317}{{\ttfamily arXiv:0912.5317
  [astro-ph.CO]}}.
[Erratum: Prog. Theor. Phys.126,351(2011)].
%%CITATION = ARXIV:0912.5317;%%.

\bibitem{Pi:2020otn}
S.~Pi and M.~Sasaki, ``{Gravitational Waves Induced by Scalar Perturbations
  with a Lognormal Peak},''
  \href{http://dx.doi.org/10.1088/1475-7516/2020/09/037}{{\em JCAP} {\bfseries
  2009} (2020) 037},
\href{http://arxiv.org/abs/2005.12306}{{\ttfamily arXiv:2005.12306 [gr-qc]}}.
%%CITATION = ARXIV:2005.12306;%%.

\bibitem{Espinosa:2018eve}
J.~R. Espinosa, D.~Racco, and A.~Riotto, ``{A Cosmological Signature of the SM
  Higgs Instability: Gravitational Waves},''
  \href{http://dx.doi.org/10.1088/1475-7516/2018/09/012}{{\em JCAP} {\bfseries
  1809} (2018) 012},
\href{http://arxiv.org/abs/1804.07732}{{\ttfamily arXiv:1804.07732 [hep-ph]}}.
%%CITATION = ARXIV:1804.07732;%%.

\bibitem{Kohri:2018awv}
K.~Kohri and T.~Terada, ``{Semianalytic calculation of gravitational wave
  spectrum nonlinearly induced from primordial curvature perturbations},''
  \href{http://dx.doi.org/10.1103/PhysRevD.97.123532}{{\em Phys. Rev.}
  {\bfseries D97} no.~12, (2018) 123532},
\href{http://arxiv.org/abs/1804.08577}{{\ttfamily arXiv:1804.08577 [gr-qc]}}.
%%CITATION = ARXIV:1804.08577;%%.

\bibitem{Braglia:2020eai}
M.~Braglia, D.~K. Hazra, F.~Finelli, G.~F. Smoot, L.~Sriramkumar, and A.~A.
  Starobinsky, ``{Generating PBHs and small-scale GWs in two-field models of
  inflation},'' \href{http://dx.doi.org/10.1088/1475-7516/2020/08/001}{{\em
  JCAP} {\bfseries 2008} (2020) 001},
\href{http://arxiv.org/abs/2005.02895}{{\ttfamily arXiv:2005.02895
  [astro-ph.CO]}}.
%%CITATION = ARXIV:2005.02895;%%.

\bibitem{Kuroyanagi:2018csn}
S.~Kuroyanagi, T.~Chiba, and T.~Takahashi, ``{Probing the Universe through the
  Stochastic Gravitational Wave Background},''
  \href{http://dx.doi.org/10.1088/1475-7516/2018/11/038}{{\em JCAP} {\bfseries
  1811} (2018) 038},
\href{http://arxiv.org/abs/1807.00786}{{\ttfamily arXiv:1807.00786
  [astro-ph.CO]}}.
%%CITATION = ARXIV:1807.00786;%%.

\bibitem{Caprini:2019pxz}
C.~Caprini, D.~G. Figueroa, R.~Flauger, G.~Nardini, M.~Peloso, M.~Pieroni,
  A.~Ricciardone, and G.~Tasinato, ``{Reconstructing the spectral shape of a
  stochastic gravitational wave background with LISA},''
  \href{http://dx.doi.org/10.1088/1475-7516/2019/11/017}{{\em JCAP} {\bfseries
  1911} (2019) 017},
\href{http://arxiv.org/abs/1906.09244}{{\ttfamily arXiv:1906.09244
  [astro-ph.CO]}}.
%%CITATION = ARXIV:1906.09244;%%.

\bibitem{Flauger:2020qyi}
R.~Flauger, N.~Karnesis, G.~Nardini, M.~Pieroni, A.~Ricciardone, and
  J.~Torrado, ``{Improved reconstruction of a stochastic gravitational wave
  background with LISA},''
  \href{http://dx.doi.org/10.1088/1475-7516/2021/01/059}{{\em JCAP} {\bfseries
  2101} (2021) 059},
\href{http://arxiv.org/abs/2009.11845}{{\ttfamily arXiv:2009.11845
  [astro-ph.CO]}}.
%%CITATION = ARXIV:2009.11845;%%.

\bibitem{Cai:2019amo}
R.-G. Cai, S.~Pi, S.-J. Wang, and X.-Y. Yang, ``{Resonant multiple peaks in the
  induced gravitational waves},''
  \href{http://dx.doi.org/10.1088/1475-7516/2019/05/013}{{\em JCAP} {\bfseries
  1905} (2019) 013},
\href{http://arxiv.org/abs/1901.10152}{{\ttfamily arXiv:1901.10152
  [astro-ph.CO]}}.
%%CITATION = ARXIV:1901.10152;%%.

\bibitem{Inomata:2019ivs}
K.~Inomata, K.~Kohri, T.~Nakama, and T.~Terada, ``{Enhancement of Gravitational
  Waves Induced by Scalar Perturbations due to a Sudden Transition from an
  Early Matter Era to the Radiation Era},''
  \href{http://dx.doi.org/10.1103/PhysRevD.100.043532}{{\em Phys. Rev.}
  {\bfseries D100} no.~4, (2019) 043532},
\href{http://arxiv.org/abs/1904.12879}{{\ttfamily arXiv:1904.12879
  [astro-ph.CO]}}.
%%CITATION = ARXIV:1904.12879;%%.

\bibitem{Fumagalli:2020nvq}
J.~Fumagalli, S.~Renaux-Petel, and L.~T. Witkowski, ``{Oscillations in the
  stochastic gravitational wave background from sharp features and particle
  production during inflation},''
\href{http://arxiv.org/abs/2012.02761}{{\ttfamily arXiv:2012.02761
  [astro-ph.CO]}}.
%%CITATION = ARXIV:2012.02761;%%.


\bibitem{Fumagalli:2021mpc}
J.~Fumagalli, G.~A.~Palma,  S.~Renaux-Petel, S.~Sypsas,   L.~T. Witkowski  and C.~Zenteno, ``{Primordial gravitational waves from excited states},''
 \href{http://dx.doi.org/10.1007/JHEP03(2022)196}{{\em JHEP}
  {\bfseries 03} (2022) 196},
\href{http://arxiv.org/abs/2111.146641}{{\ttfamily arXiv:2111.14664 [astro-ph.CO]}}.

%\href{http://arxiv.org/abs/2012.02761}{{\ttfamily arXiv:2012.02761
 % [astro-ph.CO]}}.


\bibitem{Alba:2015cms}
V.~Alba and J.~Maldacena, ``{Primordial gravity wave background
  anisotropies},'' \href{http://dx.doi.org/10.1007/JHEP03(2016)115}{{\em JHEP}
  {\bfseries 03} (2016) 115},
\href{http://arxiv.org/abs/1512.01531}{{\ttfamily arXiv:1512.01531 [hep-th]}}.
%%CITATION = ARXIV:1512.01531;%%.

\bibitem{Contaldi:2016koz}
C.~R. Contaldi, ``{Anisotropies of Gravitational Wave Backgrounds: A Line Of
  Sight Approach},''
  \href{http://dx.doi.org/10.1016/j.physletb.2017.05.020}{{\em Phys. Lett.}
  {\bfseries B771} (2017) 9--12},
\href{http://arxiv.org/abs/1609.08168}{{\ttfamily arXiv:1609.08168
  [astro-ph.CO]}}.
%%CITATION = ARXIV:1609.08168;%%.

\bibitem{Bertacca:2017vod}
D.~Bertacca, A.~Raccanelli, N.~Bartolo, and S.~Matarrese, ``{Cosmological
  perturbation effects on gravitational-wave luminosity distance estimates},''
  \href{http://dx.doi.org/10.1016/j.dark.2018.03.001}{{\em Phys. Dark Univ.}
  {\bfseries 20} (2018) 32--40},
\href{http://arxiv.org/abs/1702.01750}{{\ttfamily arXiv:1702.01750 [gr-qc]}}.
%%CITATION = ARXIV:1702.01750;%%.

\bibitem{Bartolo:2019oiq}
N.~Bartolo, D.~Bertacca, S.~Matarrese, M.~Peloso, A.~Ricciardone, A.~Riotto,
  and G.~Tasinato, ``{Anisotropies and non-Gaussianity of the Cosmological
  Gravitational Wave Background},''
  \href{http://dx.doi.org/10.1103/PhysRevD.100.121501}{{\em Phys. Rev.}
  {\bfseries D100} no.~12, (2019) 121501},
\href{http://arxiv.org/abs/1908.00527}{{\ttfamily arXiv:1908.00527
  [astro-ph.CO]}}.
%%CITATION = ARXIV:1908.00527;%%.

\bibitem{Bartolo:2019zvb}
N.~Bartolo, D.~Bertacca, V.~De~Luca, G.~Franciolini, S.~Matarrese, M.~Peloso,
  A.~Ricciardone, A.~Riotto, and G.~Tasinato, ``{Gravitational wave
  anisotropies from primordial black holes},''
  \href{http://dx.doi.org/10.1088/1475-7516/2020/02/028}{{\em JCAP} {\bfseries
  2002} (2020) 028},
\href{http://arxiv.org/abs/1909.12619}{{\ttfamily arXiv:1909.12619
  [astro-ph.CO]}}.
%%CITATION = ARXIV:1909.12619;%%.

\bibitem{Bartolo:2019yeu}
N.~Bartolo, D.~Bertacca, S.~Matarrese, M.~Peloso, A.~Ricciardone, A.~Riotto,
  and G.~Tasinato, ``{Characterizing the cosmological gravitational wave
  background: Anisotropies and non-Gaussianity},''
  \href{http://dx.doi.org/10.1103/PhysRevD.102.023527}{{\em Phys. Rev.}
  {\bfseries D102} no.~2, (2020) 023527},
\href{http://arxiv.org/abs/1912.09433}{{\ttfamily arXiv:1912.09433
  [astro-ph.CO]}}.
%%CITATION = ARXIV:1912.09433;%%.

\bibitem{Domcke:2020xmn}
V.~Domcke, R.~Jinno, and H.~Rubira, ``{Deformation of the gravitational wave
  spectrum by density perturbations},''
  \href{http://dx.doi.org/10.1088/1475-7516/2020/06/046}{{\em JCAP} {\bfseries
  2006} (2020) 046},
\href{http://arxiv.org/abs/2002.11083}{{\ttfamily arXiv:2002.11083
  [astro-ph.CO]}}.
%%CITATION = ARXIV:2002.11083;%%.

\bibitem{DallArmi:2020dar}
L.~Valbusa~Dall'Armi, A.~Ricciardone, N.~Bartolo, D.~Bertacca, and
  S.~Matarrese, ``{Imprint of relativistic particles on the anisotropies of the
  stochastic gravitational-wave background},''
  \href{http://dx.doi.org/10.1103/PhysRevD.103.023522}{{\em Phys. Rev.}
  {\bfseries D103} no.~2, (2021) 023522},
\href{http://arxiv.org/abs/2007.01215}{{\ttfamily arXiv:2007.01215
  [astro-ph.CO]}}.
%%CITATION = ARXIV:2007.01215;%%.

\bibitem{Dimastrogiovanni:2021mfs}
E.~Dimastrogiovanni, M.~Fasiello, A.~Malhotra, P.~D. Meerburg, and G.~Orlando,
  ``{Testing the Early Universe with Anisotropies of the Gravitational Wave
  Background},'' \href{http://arxiv.org/abs/2109.03077}{{\ttfamily
  arXiv:2109.03077 [astro-ph.CO]}}.

\bibitem{Ricciardone:2017kre}
A.~Ricciardone and G.~Tasinato, ``{Anisotropic tensor power spectrum at
  interferometer scales induced by tensor squeezed non-Gaussianity},''
  \href{http://dx.doi.org/10.1088/1475-7516/2018/02/011}{{\em JCAP} {\bfseries
  1802} (2018) 011},
\href{http://arxiv.org/abs/1711.02635}{{\ttfamily arXiv:1711.02635
  [astro-ph.CO]}}.
%%CITATION = ARXIV:1711.02635;%%.

\bibitem{Dimastrogiovanni:2018gkl}
E.~Dimastrogiovanni, M.~Fasiello, G.~Tasinato, and D.~Wands, ``{Tensor
  non-Gaussianities from Non-minimal Coupling to the Inflaton},''
  \href{http://dx.doi.org/10.1088/1475-7516/2019/02/008}{{\em JCAP} {\bfseries
  1902} (2019) 008},
\href{http://arxiv.org/abs/1810.08866}{{\ttfamily arXiv:1810.08866
  [astro-ph.CO]}}.
%%CITATION = ARXIV:1810.08866;%%.

\bibitem{Dimastrogiovanni:2019bfl}
E.~Dimastrogiovanni, M.~Fasiello, and G.~Tasinato, ``{Searching for Fossil
  Fields in the Gravity Sector},''
  \href{http://dx.doi.org/10.1103/PhysRevLett.124.061302}{{\em Phys. Rev.
  Lett.} {\bfseries 124} no.~6, (2020) 061302},
  \href{http://arxiv.org/abs/1906.07204}{{\ttfamily arXiv:1906.07204
  [astro-ph.CO]}}.

\bibitem{Adshead:2020bji}
P.~Adshead, N.~Afshordi, E.~Dimastrogiovanni, M.~Fasiello, E.~A. Lim, and
  G.~Tasinato, ``{Multimessenger cosmology: Correlating cosmic microwave
  background and stochastic gravitational wave background measurements},''
  \href{http://dx.doi.org/10.1103/PhysRevD.103.023532}{{\em Phys. Rev.}
  {\bfseries D103} no.~2, (2021) 023532},
\href{http://arxiv.org/abs/2004.06619}{{\ttfamily arXiv:2004.06619
  [astro-ph.CO]}}.
%%CITATION = ARXIV:2004.06619;%%.

\bibitem{Seto:2006hf}
N.~Seto, ``{Prospects for direct detection of circular polarization of
  gravitational-wave background},''
  \href{http://dx.doi.org/10.1103/PhysRevLett.97.151101}{{\em Phys. Rev. Lett.}
  {\bfseries 97} (2006) 151101},
\href{http://arxiv.org/abs/astro-ph/0609504}{{\ttfamily arXiv:astro-ph/0609504
  [astro-ph]}}.
%%CITATION = ASTRO-PH/0609504;%%.


\bibitem{Domcke:2019zls}
V.~Domcke, J.~Garcia-Bellido, M.~Peloso, M.~Pieroni, A.~Ricciardone, L.~Sorbo,
  and G.~Tasinato, ``{Measuring the net circular polarization of the stochastic
  gravitational wave background with interferometers},''
  \href{http://dx.doi.org/10.1088/1475-7516/2020/05/028}{{\em JCAP} {\bfseries
  2005} (2020) 028},
\href{http://arxiv.org/abs/1910.08052}{{\ttfamily arXiv:1910.08052
  [astro-ph.CO]}}.
%%CITATION = ARXIV:1910.08052;%%.



%\cite{Lewicki:2021kmu}
\bibitem{Lewicki:2021kmu}
M.~Lewicki and V.~Vaskonen,
``{Impact of LIGO-Virgo binaries on gravitational wave background searches},''
\href{http://arxiv.org/abs/2111.05847}{{\ttfamily arXiv:2111.05847
  [astro-ph.CO]}}.




\bibitem{Rosado:2011kv}
P.~A. Rosado, ``{Gravitational wave background from binary systems},''
  \href{http://dx.doi.org/10.1103/PhysRevD.84.084004}{{\em Phys. Rev.}
  {\bfseries D84} (2011) 084004},
\href{http://arxiv.org/abs/1106.5795}{{\ttfamily arXiv:1106.5795 [gr-qc]}}.
%%CITATION = ARXIV:1106.5795;%%.

\bibitem{Pitrou:2019rjz}
C.~Pitrou, G.~Cusin, and J.-P. Uzan, ``{Unified view of anisotropies in the
  astrophysical gravitational-wave background},''
  \href{http://dx.doi.org/10.1103/PhysRevD.101.081301}{{\em Phys. Rev. D}
  {\bfseries 101} no.~8, (2020) 081301},
  \href{http://arxiv.org/abs/1910.04645}{{\ttfamily arXiv:1910.04645
  [astro-ph.CO]}}.

\bibitem{Dvorkin:2016okx}
I.~Dvorkin, J.-P. Uzan, E.~Vangioni, and J.~Silk, ``{Synthetic model of the
  gravitational wave background from evolving binary compact objects},''
  \href{http://dx.doi.org/10.1103/PhysRevD.94.103011}{{\em Phys. Rev. D}
  {\bfseries 94} no.~10, (2016) 103011},
  \href{http://arxiv.org/abs/1607.06818}{{\ttfamily arXiv:1607.06818
  [astro-ph.HE]}}.

\bibitem{TheLIGOScientific:2017qsa}
{\bfseries LIGO Scientific, Virgo} Collaboration, B.~P. Abbott {\em et~al.},
  ``{GW170817: Observation of Gravitational Waves from a Binary Neutron Star
  Inspiral},'' \href{http://dx.doi.org/10.1103/PhysRevLett.119.161101}{{\em
  Phys. Rev. Lett.} {\bfseries 119} no.~16, (2017) 161101},
  \href{http://arxiv.org/abs/1710.05832}{{\ttfamily arXiv:1710.05832 [gr-qc]}}.

\bibitem{LIGOScientific:2018mvr}
{\bfseries LIGO Scientific, Virgo} Collaboration, B.~P. Abbott {\em et~al.},
  ``{GWTC-1: A Gravitational-Wave Transient Catalog of Compact Binary Mergers
  Observed by LIGO and Virgo during the First and Second Observing Runs},''
  \href{http://dx.doi.org/10.1103/PhysRevX.9.031040}{{\em Phys. Rev. X}
  {\bfseries 9} no.~3, (2019) 031040},
  \href{http://arxiv.org/abs/1811.12907}{{\ttfamily arXiv:1811.12907
  [astro-ph.HE]}}.

\bibitem{LIGOScientific:2019vic}
{\bfseries LIGO Scientific, Virgo} Collaboration, B.~P. Abbott {\em et~al.},
  ``{Search for the isotropic stochastic background using data from Advanced
  LIGOs second observing run},''
  \href{http://dx.doi.org/10.1103/PhysRevD.100.061101}{{\em Phys. Rev.}
  {\bfseries D100} no.~6, (2019) 061101},
\href{http://arxiv.org/abs/1903.02886}{{\ttfamily arXiv:1903.02886 [gr-qc]}}.
%%CITATION = ARXIV:1903.02886;%%.


\bibitem{Cusin:2019jpv}
G.~Cusin, I.~Dvorkin, C.~Pitrou, and J.-P. Uzan, ``{Properties of the
  stochastic astrophysical gravitational wave background: astrophysical sources
  dependencies},'' \href{http://dx.doi.org/10.1103/PhysRevD.100.063004}{{\em
  Phys. Rev. D} {\bfseries 100} no.~6, (2019) 063004},
  \href{http://arxiv.org/abs/1904.07797}{{\ttfamily arXiv:1904.07797
  [astro-ph.CO]}}.

\bibitem{Cusin:2019jhg}
G.~Cusin, I.~Dvorkin, C.~Pitrou, and J.-P. Uzan, ``{Stochastic gravitational
  wave background anisotropies in the mHz band: astrophysical dependencies},''
  \href{http://dx.doi.org/10.1093/mnrasl/slz182}{{\em Mon. Not. Roy. Astron.
  Soc.} {\bfseries 493} no.~1, (2020) L1--L5},
  \href{http://arxiv.org/abs/1904.07757}{{\ttfamily arXiv:1904.07757
  [astro-ph.CO]}}.

\bibitem{Cusin:2018rsq}
G.~Cusin, I.~Dvorkin, C.~Pitrou, and J.-P. Uzan, ``{First predictions of the
  angular power spectrum of the astrophysical gravitational wave background},''
  \href{http://dx.doi.org/10.1103/PhysRevLett.120.231101}{{\em Phys. Rev.
  Lett.} {\bfseries 120} (2018) 231101},
\href{http://arxiv.org/abs/1803.03236}{{\ttfamily arXiv:1803.03236
  [astro-ph.CO]}}.
%%CITATION = ARXIV:1803.03236;%%.

\bibitem{Jenkins:2018uac}
A.~C. Jenkins, M.~Sakellariadou, T.~Regimbau, and E.~Slezak, ``{Anisotropies in
  the astrophysical gravitational-wave background: Predictions for the
  detection of compact binaries by LIGO and Virgo},''
  \href{http://dx.doi.org/10.1103/PhysRevD.98.063501}{{\em Phys. Rev.}
  {\bfseries D98} no.~6, (2018) 063501},
\href{http://arxiv.org/abs/1806.01718}{{\ttfamily arXiv:1806.01718
  [astro-ph.CO]}}.
%%CITATION = ARXIV:1806.01718;%%.

\bibitem{Cusin:2017mjm}
G.~Cusin, C.~Pitrou, and J.-P. Uzan, ``{The signal of the gravitational wave
  background and the angular correlation of its energy density},''
  \href{http://dx.doi.org/10.1103/PhysRevD.97.123527}{{\em Phys. Rev.}
  {\bfseries D97} no.~12, (2018) 123527},
\href{http://arxiv.org/abs/1711.11345}{{\ttfamily arXiv:1711.11345
  [astro-ph.CO]}}.
%%CITATION = ARXIV:1711.11345;%%.

\bibitem{Cusin:2017fwz}
G.~Cusin, C.~Pitrou, and J.-P. Uzan, ``{Anisotropy of the astrophysical
  gravitational wave background: Analytic expression of the angular power
  spectrum and correlation with cosmological observations},''
  \href{http://dx.doi.org/10.1103/PhysRevD.96.103019}{{\em Phys. Rev.}
  {\bfseries D96} no.~10, (2017) 103019},
\href{http://arxiv.org/abs/1704.06184}{{\ttfamily arXiv:1704.06184
  [astro-ph.CO]}}.
%%CITATION = ARXIV:1704.06184;%%.

\bibitem{Jenkins:2019uzp}
A.~C. Jenkins and M.~Sakellariadou, ``{Shot noise in the astrophysical
  gravitational-wave background},''
\href{http://arxiv.org/abs/1902.07719}{{\ttfamily arXiv:1902.07719
  [astro-ph.CO]}}.
%%CITATION = ARXIV:1902.07719;%%.

\bibitem{Jenkins:2019nks}
A.~C. Jenkins, J.~D. Romano, and M.~Sakellariadou, ``{Estimating the angular
  power spectrum of the gravitational-wave background in the presence of shot
  noise},''
\href{http://arxiv.org/abs/1907.06642}{{\ttfamily arXiv:1907.06642
  [astro-ph.CO]}}.
%%CITATION = ARXIV:1907.06642;%%.

\bibitem{Cusin:2018avf}
G.~Cusin, R.~Durrer, and P.~G. Ferreira, ``{Polarization of a stochastic
  gravitational wave background through diffusion by massive structures},''
  \href{http://dx.doi.org/10.1103/PhysRevD.99.023534}{{\em Phys. Rev. D}
  {\bfseries 99} no.~2, (2019) 023534},
  \href{http://arxiv.org/abs/1807.10620}{{\ttfamily arXiv:1807.10620
  [astro-ph.CO]}}.

\bibitem{Bertacca:2019fnt}
D.~Bertacca, A.~Ricciardone, N.~Bellomo, A.~C. Jenkins, S.~Matarrese,
  A.~Raccanelli, T.~Regimbau, and M.~Sakellariadou, ``{Projection effects on
  the observed angular spectrum of the astrophysical stochastic gravitational
  wave background},'' \href{http://dx.doi.org/10.1103/PhysRevD.101.103513}{{\em
  Phys. Rev.} {\bfseries D101} no.~10, (2020) 103513},
\href{http://arxiv.org/abs/1909.11627}{{\ttfamily arXiv:1909.11627
  [astro-ph.CO]}}.
%%CITATION = ARXIV:1909.11627;%%.

%\cite{Mukherjee:2019oma}
\bibitem{Mukherjee:2019oma}
S.~Mukherjee and J.~Silk,
``{Time-dependence of the astrophysical stochastic gravitational wave background},''
%Mon. Not. Roy. Astron. Soc. \textbf{491} (2020) no.4, 4690-4701
%doi:10.1093/mnras/stz3226
\href{http://arxiv.org/abs/1912.07657}{{\ttfamily arXiv:1912.07657
  [gr-qc]}}.
  %
%[arXiv:1912.07657 [gr-qc]].
%25 citations counted in INSPIRE as of 02 Feb 2022

\bibitem{Alonso:2020mva}
D.~Alonso, G.~Cusin, P.~G. Ferreira, and C.~Pitrou, ``{Detecting the
  anisotropic astrophysical gravitational wave background in the presence of
  shot noise through cross-correlations},''
  \href{http://dx.doi.org/10.1103/PhysRevD.102.023002}{{\em Phys. Rev. D}
  {\bfseries 102} no.~2, (2020) 023002},
  \href{http://arxiv.org/abs/2002.02888}{{\ttfamily arXiv:2002.02888
  [astro-ph.CO]}}.

\bibitem{LIGOScientific:2019gaw}
{\bfseries LIGO Scientific, Virgo} Collaboration, B.~P. Abbott {\em et~al.},
  ``{Directional limits on persistent gravitational waves using data from
  Advanced LIGO's first two observing runs},''
  \href{http://dx.doi.org/10.1103/PhysRevD.100.062001}{{\em Phys. Rev. D}
  {\bfseries 100} no.~6, (2019) 062001},
  \href{http://arxiv.org/abs/1903.08844}{{\ttfamily arXiv:1903.08844 [gr-qc]}}.

\bibitem{Baker:2019ync}
J.~Baker {\em et~al.}, ``{High angular resolution gravitational wave
  astronomy},''
\href{http://arxiv.org/abs/1908.11410}{{\ttfamily arXiv:1908.11410
  [astro-ph.HE]}}.
%%CITATION = ARXIV:1908.11410;%%.

\bibitem{Yang:2020usq}
K.~Z. Yang, V.~Mandic, C.~Scarlata, and S.~Banagiri, ``{Searching for
  Cross-Correlation Between Stochastic Gravitational Wave Background and Galaxy
  Number Counts},'' \href{http://dx.doi.org/10.1093/mnras/staa3159}{{\em Mon.
  Not. Roy. Astron. Soc.} {\bfseries 500} no.~2, (2020) 1666--1672},
  \href{http://arxiv.org/abs/2007.10456}{{\ttfamily arXiv:2007.10456
  [astro-ph.CO]}}.

\bibitem{Lamberts:2018cub}
A.~Lamberts, S.~Garrison-Kimmel, P.~Hopkins, E.~Quataert, J.~Bullock, C.-A.
  Faucher-Giguere, A.~Wetzel, D.~Keres, K.~Drango, and R.~Sanderson,
  ``{Predicting the binary black hole population of the Milky Way with
  cosmological simulations},''
  \href{http://dx.doi.org/10.1093/mnras/sty2035}{{\em Mon. Not. Roy. Astron.
  Soc.} {\bfseries 480} no.~2, (2018) 2704--2718},
\href{http://arxiv.org/abs/1801.03099}{{\ttfamily arXiv:1801.03099
  [astro-ph.GA]}}.
%%CITATION = ARXIV:1801.03099;%%.

\bibitem{Lamberts:2019nyk}
A.~Lamberts, S.~Blunt, T.~B. Littenberg, S.~Garrison-Kimmel, T.~Kupfer, and
  R.~E. Sanderson, ``{Predicting the LISA white dwarf binary population in the
  Milky Way with cosmological simulations},''
  \href{http://dx.doi.org/10.1093/mnras/stz2834}{{\em Mon. Not. Roy. Astron.
  Soc.} {\bfseries 490} no.~4, (2019) 5888--5903},
\href{http://arxiv.org/abs/1907.00014}{{\ttfamily arXiv:1907.00014
  [astro-ph.HE]}}.
%%CITATION = ARXIV:1907.00014;%%.

\bibitem{Robson:2017ayy}
T.~Robson, N.~Cornish,  ``{Impact of galactic foreground characterization on a global analysis for the LISA gravitational wave observatory},'' {{\em Class. Quant. Grav.} {\bfseries 34} (2017)  244002}, \href{http://arxiv.org/abs/1705.09421}{{\ttfamily arXiv:1705.09421
  [gr-qc]}}.

\bibitem{Dodelson:2003ft}
S.~Dodelson, {\em {Modern Cosmology}}.
\newblock Academic Press, Amsterdam, 2003.
\newblock
\url{http://www.slac.stanford.edu/spires/find/books/www?cl=QB981:D62:2003}.
\newblock
%%CITATION = INSPIRE-640063;%%.

\bibitem{Cusin:2016kqx}
G.~Cusin, C.~Pitrou, and J.-P. Uzan, ``{Are we living near the center of a
  local void?},'' \href{http://dx.doi.org/10.1088/1475-7516/2017/03/038}{{\em
  JCAP} {\bfseries 1703} no.~03, (2017) 038},
\href{http://arxiv.org/abs/1609.02061}{{\ttfamily arXiv:1609.02061
  [astro-ph.CO]}}.
%%CITATION = ARXIV:1609.02061;%%.

\bibitem{Nistane:2019yzd}
V.~Nistane, G.~Cusin, and M.~Kunz, ``{CMB sky for an off-center observer in a
  local void. Part I. Framework for forecasts},''
  \href{http://dx.doi.org/10.1088/1475-7516/2019/12/038}{{\em JCAP} {\bfseries
  12} (2019) 038}, \href{http://arxiv.org/abs/1908.05484}{{\ttfamily
  arXiv:1908.05484 [astro-ph.CO]}}.

\bibitem{Alonso:2020rar}
D.~Alonso, C.~R. Contaldi, G.~Cusin, P.~G. Ferreira, and A.~I. Renzini,
  ``{Noise angular power spectrum of gravitational wave background
  experiments},'' \href{http://dx.doi.org/10.1103/PhysRevD.101.124048}{{\em
  Phys. Rev. D} {\bfseries 101} no.~12, (2020) 124048},
  \href{http://arxiv.org/abs/2005.03001}{{\ttfamily arXiv:2005.03001
  [astro-ph.CO]}}.

\bibitem{Reitze:2019iox}
D.~Reitze {\em et~al.}, ``{Cosmic Explorer: The U.S. Contribution to
  Gravitational-Wave Astronomy beyond LIGO},'' {\em Bull. Am. Astron. Soc.}
  {\bfseries 51} no.~7, (2019) 035,
\href{http://arxiv.org/abs/1907.04833}{{\ttfamily arXiv:1907.04833
  [astro-ph.IM]}}.
%%CITATION = ARXIV:1907.04833;%%.

\bibitem{hild2008pushing}
S.~Hild, S.~Chelkowski, and A.~Freise, ``Pushing towards the et sensitivity
  using 'conventional' technology,'' 2008.

\bibitem{Dalang:2021ruy}
C.~Dalang, C.~Bonvin, ``{On the kinematic cosmic dipole tension}'',
 \href{http://arxiv.org/abs/2111.0361}{{\ttfamily arXiv:2111.0361
  [astro-ph.CO]}}.



%{\small
%\addcontentsline{toc}{section}{References}
%\bibliographystyle{utphys}

%%\bibliographystyle{utcaps}e
%%\bibliographystyle{kp}


%\bibliography{DIPOLErefs}
%}
\end{thebibliography}
\end{document}